\documentclass[11pt]{article}

\usepackage{amsmath,amsfonts,amssymb,amsthm}
\usepackage[margin=1in]{geometry}
\usepackage[colorlinks]{hyperref}
\usepackage{graphicx}
\usepackage{algorithm}
\usepackage{algorithmicx}
\usepackage{authblk}
\usepackage{algpseudocode}
\usepackage{bbm}
\usepackage{comment}
\usepackage{accents}
\usepackage{adjustbox}
\usepackage{booktabs}
\usepackage{caption}
\usepackage{subcaption} 
\usepackage{tikz}
\usepackage{overpic}

\usepackage{bm}

\theoremstyle{definition}
\newtheorem{theorem}{Theorem}

\newtheorem{definition}{Definition}[section]

\newcommand{\Be}{\bm{e}}

\newcommand{\Bp}{\bm{p}}

\def\bF{\mathbf{F}}

\def\bR{\mathbf{R}}

\def\bG{\mathbf{G}}

\def\bR{\mathbb{R}}

\def\bF{\mathbf{F}}

\def\bx{\mathbf{x}}

\newcommand{\quotes}[1]{``#1''}

\begin{document}

\title{A Fast Algorithm for the Finite Expression Method in Learning Dynamics on Complex Networks}

\author[1]{Zezheng Song\thanks{Email: \texttt{zsong001@umd.edu}.}}
\author[2]{Chunmei Wang\thanks{Email: \texttt{chunmei.wang@ufl.edu}.}}
\author[1,3]{Haizhao Yang\thanks{Email: \texttt{hzyang@umd.edu}.} }

\affil[1]{Department of Mathematics, University of Maryland, College Park, MD 20742, USA}
\affil[2]{Department of Mathematics, University of Florida, Gainesville, FL 32611, USA   }
\affil[3]{Department of Computer Science, University of Maryland, College Park, MD 20742, USA}

\date{}
\maketitle

\begin{abstract}
    Complex network data is prevalent in various real-world domains, including physical, technological, and biological systems. Despite this prevalence, predicting trends and understanding behavioral patterns in complex systems remain challenging due to poorly understood underlying mechanisms. While data-driven methods have advanced in uncovering governing equations from time series data, efforts to extract physical laws from network data are limited and often struggle with incomplete or noisy data. Additionally, they suffer from computational costs on network data, making it difficult to scale to real-world networks. To address these challenges, we introduce a novel approach called the Finite Expression Method (FEX) and its fast algorithm for learning dynamics on complex networks. FEX represents dynamics on complex networks using binary trees composed of finite mathematical operators. The nodes within these trees are trained through a combinatorial optimization process guided by reinforcement learning techniques. This unique configuration allows FEX to capture complex dynamics with minimal prior knowledge of the system and a small dictionary of mathematical operators. We also integrate a fast, stochastic algorithm into FEX, reducing the computational complexity from $O(N^2)$ to $O(N)$. Our extensive numerical experiments demonstrate that FEX excels in accurately identifying dynamics across diverse network topologies and dynamic behaviors.
\end{abstract}


\maketitle
\section{Introduction}\label{sec:intro}

Complex network data is prevalent in various real-world domains, spanning from transportation~\cite{lin2013complex,soh2010weighted,xu2008exploring,zanin2018studying} to biology~\cite{cho2012chapter,bracken2016network,lima2009powerful} and epidemiology~\cite{lloyd2007network,danon2011networks,pastor2015epidemic}.   For instance, proteins frequently interact with one another to execute a wide range of functions. As with many complex systems, comprehending the underlying dynamics within these networks is crucial for analyzing and predicting their behaviors. In practice, time-series data for individual nodes within a network are often accessible. Still, extracting the dynamic interactions among these nodes to gain insights into the system's behavior proves to be a challenging endeavor. Additionally,
the interaction dynamics among particles in the network significantly increase the computational expense of identifying physical laws. Consequently, the physical laws governing the majority of real-world complex networks remain elusive and underexplored.

In recent years, machine learning has emerged as a promising avenue for data-driven discovery of physical laws. Notably, deep learning has demonstrated its potential to identify governing partial differential equations (PDEs) from noisy data ~\cite{chen1990non,gonzalez1998identification, long2019pde,lusch2018deep,liu2023prose,sun2020neupde,karniadakis2021physics,kemeth2023black,dietrich2023learning,psarellis2022data,arbabi2020linking,bertalan2019learning,gonzalez1998identification,churchill2022learning,churchill2023flow,qin2019data,fu2020learning,lagergren2020biologically,vlachas2018data,harlim2021machine,bongard2007automated,messenger2021weak,rudy2017data,schaeffer2017learning,champion2019data, chen2024learning, qin2021data,schmidt2009distilling,chang2019identification, long2018pde}. The success of these methods is attributed to the expressive power of deep learning models in capturing relationships among different variables. However, deep learning methods suffer from the black-box nature of their solutions, making it difficult to interpret and explicitly identify the governing equation in mathematical form. To address this, recent works have explored symbolic regression techniques to infer governing equations, thereby obtaining an explicit mathematical representation of the system~\cite{udrescu2020ai, landajuela2022unified, sun2022symbolic, iba1995numerical, gray1998nonlinear}. Nonetheless, there has been limited focus on the discovery of physical laws within complex networks. Discovering dynamics on network data involves various challenges:
\begin{itemize}
\item\textbf{Complexity and scale}: Real-world network data can be extremely large and complex, making analysis computationally expensive. In the full interaction model with $N$ nodes in the network, each node interacts with every other particle, leading to $N(N-1)$ interaction terms that need to be computed at each time, resulting a $O(N^2)$ complexity. 

\item \textbf{Data quality and availability}: The quality of network data can vary greatly. Issues like missing data, noise, and topological inaccuracies can significantly impact the analysis.
\end{itemize}

In Subsection~\ref{sec:related_work}, we summarize some of the recent work on learning dynamics from network data, and we also introduce some of the recent progress on fast algorithms. In Subsection~\ref{sec:contributions}, we introduce the overall idea of our proposed method and how it tackles these challenges.


 \subsection{Related work} \label{sec:related_work}

 In this Subsection, we introduce related work on identifying governing equations on network data and fast algorithms for scientific computing.

\subsubsection{Discovering physical laws on network}

\textbf{SINDy}: The Sparse Identification of Nonlinear Dynamics (SINDy)~\cite{brunton2016discovering} algorithm aims to extract underlying governing equations from data in a interpretable and efficient way, with theoretical convergence analysis~\cite{zhang2019convergence}. Taking time-series data as inputs, SINDy infers the underlying physical laws of the system by using Sequentially Thresholded Least Squares algorithm to choose a parsimonious model from a large library of candidate functions, usually including polynomials and trignometric functions. SINDy has been used to identify models in fluid dynamics~\cite{fukami2021sparse}, control systems~\cite{kaiser2018sparse}, and neuroscience~\cite{daniels2015automated}, etc. However, it has been shown that the vanilla SINDy performs poorly in identifying dynamics on network data. In addition, SINDy assumes direct access to the time derivative of the dynamics in the candidate function library, which is often unavailable in most real-world applications.

\textbf{Two-Phase SINDy}: In \cite{gao2022autonomous}, Gao and Yan proposed a two-phase procedure consisting of global regression and local fine-tuning for graph dynamics inference. This method works by building two comprehensive function libraries, $L_F$ and $L_G$, for self and interaction dynamics. Then, in the first phase, this method performs a sparse regression on the normalized input data and the libraries $L_F$ and $L_G$ to identify potential terms in the network dynamics. In the second phase, it performs topological sampling to fine-tune the coefficients of functions from the first phase. This method can identify the explicit functional forms of the system dynamics, given sufficiently large function libraries. However, users often lack a priori knowledge of the candidate functions, so they assume a prior solid knowledge of the system. In addition, this method is sensitive to noisy and low-resolution data because topological samplings are used in the fine-tuning stage.

\textbf{Graph Neural Networks}: Graph neural networks (GNNs) are popular models to learn the dynamics on the networks and make predictions on the systems~\cite{murphy2021deep,zang2020neural,fritz2022combining,gao2021stan}. These models typically use an encoder neural network to denote node states and create a latent representation. Following this, Neural Ordinary Differential Equations (Neural ODEs) are applied on the latent space to learn the dynamics. Finally, a decoder is used to project the latent embedding back to the original high-dimensional state. While this approach is flexible enough to be applied to a wide variety of network data, it essentially functions as a ``black box", making it difficult to understand the underlying dynamics. Additionally, it struggles to generalize well to out-of-distribution data.

 \textbf{ARNI}: Algorithm for Revealing Network Interactions (ARNI)~\cite{casadiego2017model} is a model-free method originally proposed to identify direct interactions of particles in a network data. It works by decomposing the dynamics of each node as a linear combination of basis functions in pairwise and higher-order interactions with other nodes in the system, and recast the reconstruction problem into a mathematical regression problem with grouped variables. Authors of~\cite{gao2022autonomous} modified ARNI appropriately to make it capable of identifying network dynamics instead of only inferring the graph topology.

 \subsubsection{Fast algorithms for scientific computation}
In the realm of physical sciences, it is important to develop fast and efficient algorithms to handle complex systems while keeping a high accuracy. For example, Jin et al.~\cite{jin2020random} proposed the Random Batch Methods (RBM), which splits particles in a physical system into several batches at each time to reduce the computational cost from $O(N^2)$ to $O(N)$, where $N$ is the number of particles. In machine learning, stochastic gradient descent (SGD)~\cite{amari1993backpropagation,bottou2010large,hardt2016train} takes a random sample of the entire dataset in each iteration to compute the gradient and update model parameters. Similar to SGD, stochastic coordinate descent~\cite{liu2015asynchronous,liu2014asynchronous} randomly samples some of the coordinates to update at each time step. In addition, randomized algorithms have been widely applied to fast matrix factorizations~\cite{drineas2006fast,halko2011finding,martinsson2011randomized} and Bayesian inference~\cite{welling2011bayesian,ma2015complete}, etc. In spirit, our work is similar to RBM; however, instead of taking a random batch of particles for the evolution of the dynamics, we use a stochastic approach to select a random batch of particles at each step for discovering the physical laws, i.e., solving an optimization problem.

\subsection{Our contributions} \label{sec:contributions}

\begin{figure}[ht]
    \centering
    \includegraphics[width=0.8\linewidth]{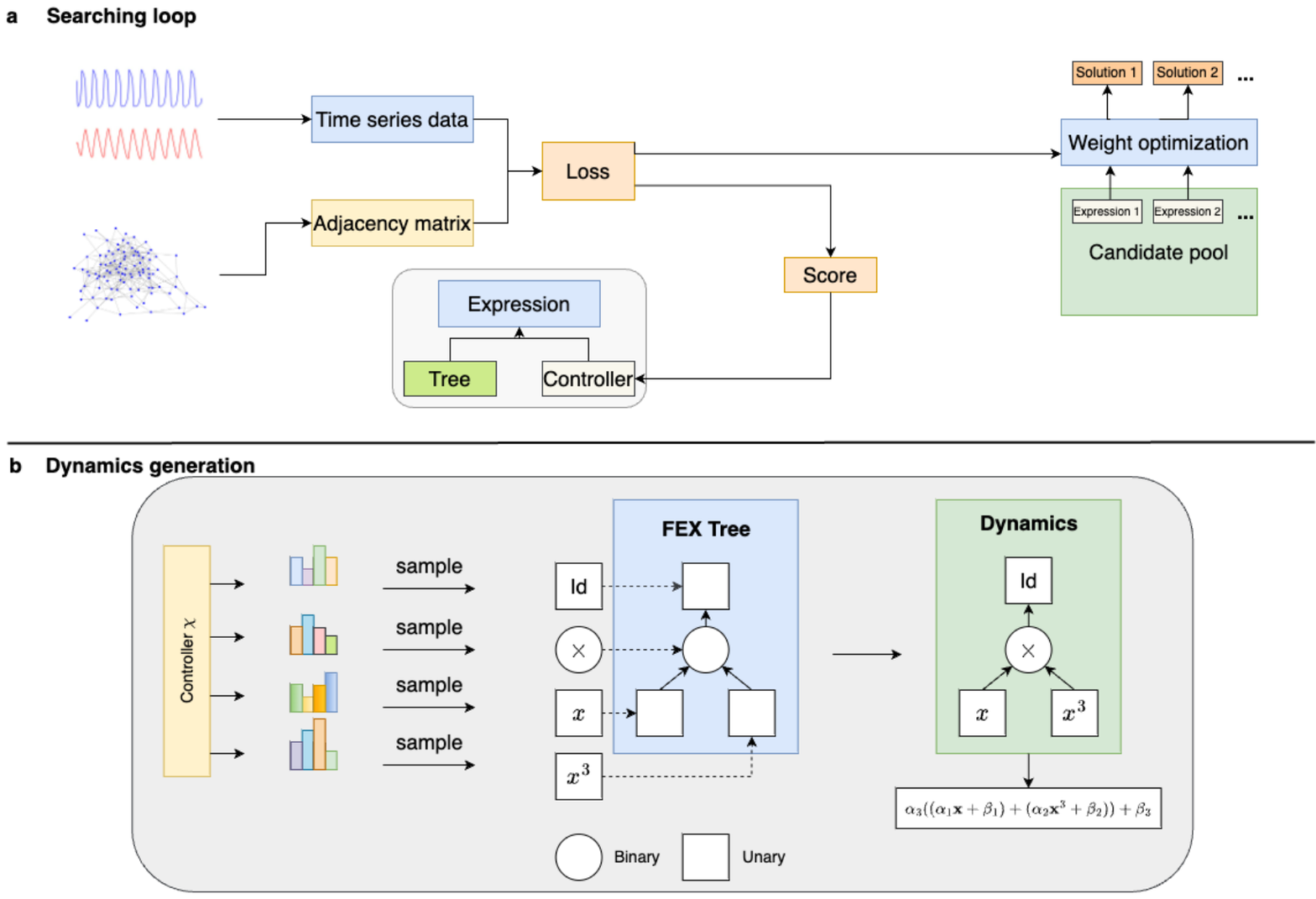}
    \caption{(A) Based on the input data of the time series and adjacency matrix, FEX performs combinatorial optimization to infer the mathematical structures of the dynamics and then implements fine-tuning on the top-performing candidates in the pool to identify the optimal dynamics. (B) FEX implements combinatorial optimization by training a controller $\chi$ to output probability mass function for each node in the FEX binary tree.}

    \label{fig:overview}
\end{figure}

To tackle the problems seen in previous methods, we propose the finite expression method (FEX) that represents dynamics on network data by binary trees, whose nodes are mathematical operators (e.g., sin, cos, exp, etc). FEX selects the operator of each node via combinatorial optimization (CO) by a novel reinforcement learning (RL) approach. In particular, our method has the following advantages: (1) Compared to the widely used SINDy approach and its variants, FEX does not require a large library of candidate functions. In contrast, thanks to the binary tree representation, FEX can produce complicated composition of functions with a few simple mathematical operators provided by the user. In addition, unlike SINDy, FEX directly estimates the time derivatives from time series data without direct access to the derivatives. This is a more challenging task due to the difficulty of estimating derivatives from noisy time series data. (2) Our model employs a RL approach to perform operator selection in a data-driven manner (discussed in Section~\ref{sec:methods}). This design has been demonstrated to promote the expressiveness of our method of identifying network dynamics. (3) To mitigate the intensive computational burden when calculating the pair-wise interaction dynamics on network data, we utilize a stochastic algorithm
to accelerate the computation while keeping a reasonable accuracy. Throughout this work, we demonstrate the advantage of our method by applying it to several synthetic complex network data to extract governing equations. An overview figure of FEX to infer physical laws on network data is provided in Fig.~\ref{fig:overview}. 

\section{Problem setup}\label{sec:background}

\subsection{Learning dynamics on complex networks}

A common model~\cite{gao2022autonomous,liu2023we} that describes the evolution of the states of the network of size $N$ is given by
\begin{equation}
\label{eqn:network_dynamics}
    \frac{\mathrm{d} \mathbf{x}_i(t)}{\mathrm{d} t}=\mathbf{F}\left(\mathbf{x}_i(t)\right)+\sum_{j=1}^N \mathbf{A}_{i j} \mathbf{G}\left(\mathbf{x}_i(t), \mathbf{x}_j(t)\right),
\end{equation}
where $\mathbf{x}_i(t) \equiv\left(x_{i, 1}(t), \ldots, x_{i, d}(t)\right)^{\top}$  represents node $i$'s $d$-dimensional feature and $i = 1,\ldots, N$. The dynamics of $\mathbf{x}_i(t)$ are driven by $\mathbf{F}(\mathbf{x}_i) \equiv \left(F_1\left(\mathbf{x}_i\right), \ldots, F_d\left(\mathbf{x}_i\right)\right)^{\top}$, denoting the self-dynamics, and $\mathbf{G}(\mathbf{x}_i(t),\mathbf{x}_j(t))\equiv \left(G_1\left(\mathbf{x}_i, \mathbf{x}_j\right), \ldots, G_d\left(\mathbf{x}_i, \mathbf{x}_j\right)\right)^{\top}$, representing the interaction dynamics between any pair of nodes $\mathbf{x}_i$ and $\mathbf{x}_j$. $\mathbf{A}_{ij}$ denotes the $ij$-th entry in the $N \times N$ adjacency matrix $\mathbf{A}$, where $\mathbf{A}_{ij} = 1$ if there is a connection between node $i$ and node $j$, and $\mathbf{A}_{ij} = 0$ otherwise. Therefore, the dynamics are determined by self-dynamics, interaction dynamics and the topology of the network. However, in many real-world complex network systems, we do not have the information of the explicit forms of $\mathbf{F}$ and $\mathbf{G}$. Therefore, it is important to distill the functional forms of $\mathbf{F}$ and $\mathbf{G}$ from node activity data, which is the focus of our work.

However, inferring underlying dynamics from real-world time series data presents a formidable challenge, primarily due to the often unclear or only partially known governing laws. The endeavor to discover dynamics within a complex network system introduces an additional layer of complexity, attributable to the interactions among entities described by the graph topology. Moreover, numerous oscillator systems exhibit varied coupling behaviors contingent upon the coupling strength between entities and the inherent graph topology, a prime example being synchronization phenomena. In addition, the data in real-world are often sparse, noisy, and come from incomplete network topology. Hence, to demonstrate the capability of FEX to discover dynamics on complex networks, we subject FEX to a rigorous examination against three classical network dynamics: Hindmarsh-Rose (HR, $d=3$)~\cite{hindmarsh1982model, storace2008hindmarsh, wang1993genesis}, FitzHugh-Nagumo (FHN, $d=2$) neuronal systems~\cite{fitzhugh1961impulses}, and coupled heterogeneous R\"ossler oscillators ($d=3$)~\cite{barzel2015constructing}, where $d$ denotes the dimension of feature for each node. Regarding to the graph topology, we demonstrate that FEX is capable of identifying governing equations on various synthetic networks, including Erd\H{o}s-R\'enyi (ER)~\cite{gomez2006scale} and scale-free (SF)~\cite{albert2002statistical} networks. Through this extensive examination, we aim to demonstrate the effectiveness and robustness of FEX in decoding the intricate dynamics inherent in complex network systems. Details are elaborated more in Section~\ref{sec:results}.

\subsection{FEX for learning dynamics on networks}

\begin{figure}[ht]
    \centering
    \includegraphics[width=0.8\linewidth]{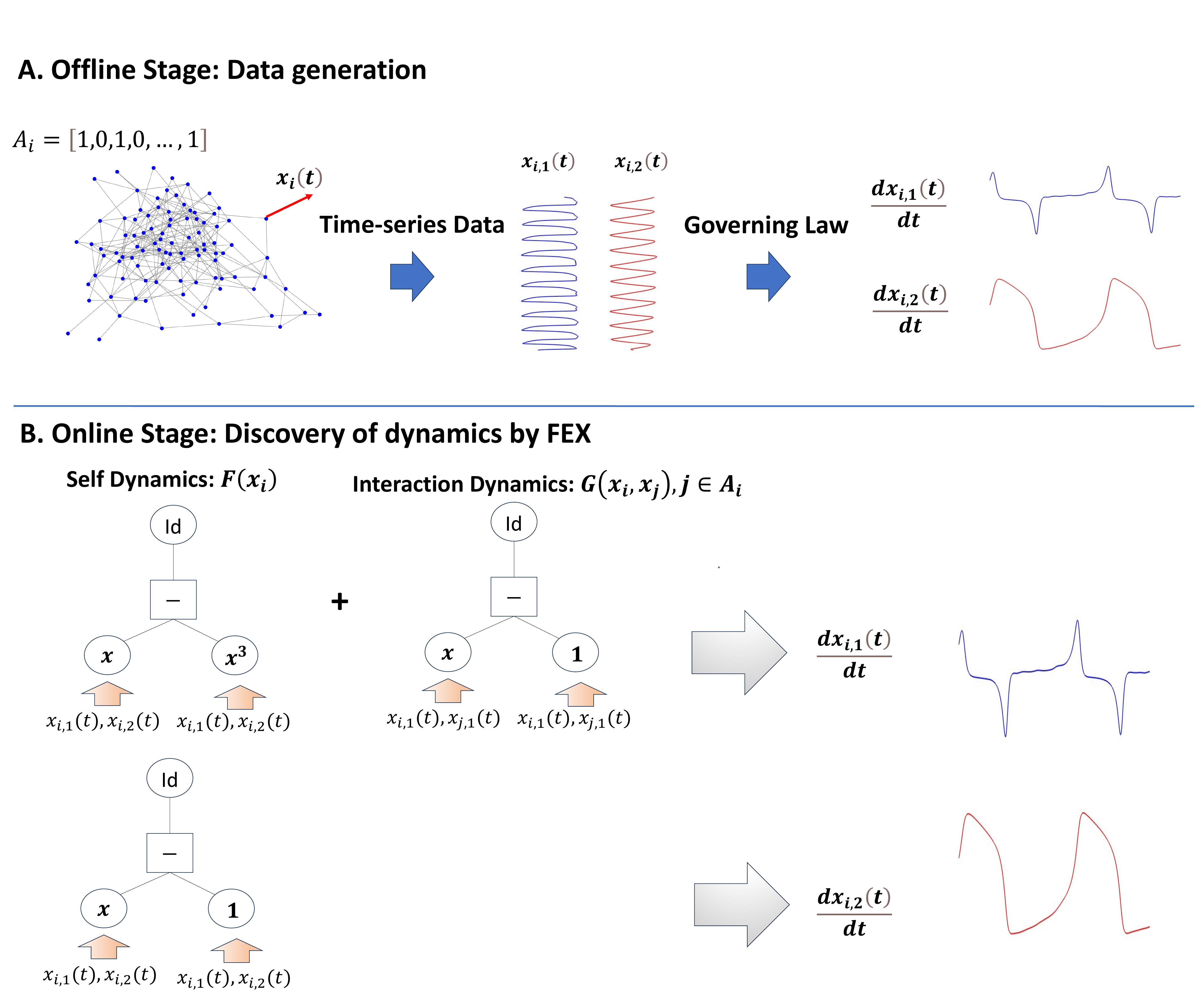}
    \caption{Representation of the components of our implementation with FHN dynamics. (A) In the offline stage, we generate time series data according to the governing law of the system, potentially with noise or low resolution. (B) In the online phase, based on the time series data, we use two FEX binary trees to infer the dynamics of the system. }

    \label{fig:workflow}
\end{figure}

The motivation behind FEX is to develop a robust methodology for identifying physical laws from network data. Our model depicts network dynamics using  two fixed-size binary trees: one representing the self-dynamics $\bF(\bx_i)$ while another one representing the interaction dynamics $\bG(\bx_i, \bx_j)$. Each tree node denotes a simple mathematical operator—either unary (sin, cos, and exp ...) or binary ($+, -, \times$...). The aim of the learning process is to identify the most suitable mathematical operator for each node of the tree, ensuring accurate representation of the dynamics. The input to FEX algorithm consists of time series data $\bx_i$ for each node in the network, and the adjacency matrix $\mathbf{A}$  representing the network topology, as illustrated by Fig.~\ref{fig:workflow}.

\subsection{Learning dynamics from sparse and noisy data}
In real-world scenarios, data are inevitably contaminated by noise or by the incompleteness of network topology. Thus, evaluating the robustness of the proposed method across diverse scenarios—such as low resolution, observational noises, and spurious or missing links—is paramount. Moreover, a comparative analysis with baseline models is conducted to ascertain the relative efficacy of the proposed method.

\textbf{Low resolution}: Practical constraints or technological limitations often result in the collection of sparse raw data. To emulate such circumstances, we simulate nonlinear dynamics with a time step size of 
$t=0.01$, followed by a uniform down-sampling of the time series data. The outcomes reveal that FEX retains a high degree of accuracy, i.e., relative errors of coefficients identified by FEX and the true coefficients are less than $1\%$, even with  merely
$5\%$ or $10\%$ of the original data.

\textbf{Spurious and missing links}: In real-world applications, the evolving network topology makes capturing all node interactions a challenging task. To assess the robustness of FEX against incomplete topology, we randomly either add or remove a proportion of links from the authentic network topology. The results indicate that FEX can accommodate approximately 
$15\%$ missing or spurious links, showcasing its robustness in handling topological inaccuracies.

\textbf{Observational noise}: Observational noise is another common issue in data measurement. We replicate such noise by introducing Gaussian noises to the time series data for every node within the network. The noise intensity is modulated via the signal-to-noise ratio (SNR). The findings underscore that FEX sustains a high accuracy level, even amidst a 40-dB observational noise environment. In contrast, other baseline methods such as Two-Phase SINDy~\cite{gao2022autonomous} and ARNI~\cite{casadiego2017model} identified erroneous and redundant terms with low-accuracy coefficients.

Through this thorough evaluation across varied scenarios and a comparative analysis with baseline models, we aim to demonstrate the robustness and stability of FEX in extracting dynamics from real-world, noisy data with potential topological inaccuracies. Further details are provided in Supplementary Information.

\section{Methods}\label{sec:methods}

\subsection{Overview of FEX}

FEX~\cite{liang2022finite, song2025finite} introduces a novel strategy for solving PDEs and identifying physical laws from data~\cite{jiang2023finite}. FEX represents the system's dynamics using a learnable binary tree to identify dynamics on network data. This tree comprises unary operators (e.g., sin, cos, etc.) and binary operators ($+, - , \times$, etc.). Additionally, each unary operator incorporates learnable weights and biases, augmenting the expressiveness of the expression. Given a specific set of operators in the tree, input variables $\bx \in \bR^d$ are passed through the FEX tree via the leaves. The complete FEX expression is then derived through a preorder traversal of the tree. Unlike many symbolic regression (SR) methods that generate equations autoregressively, FEX allows users to predetermine the tree's depth and size. This distinctive feature has been demonstrated to enhance FEX's accuracy in solving PDEs and discovering governing equations. Subsequently, with a suitable loss function, FEX conducts CO using a RL strategy to perform operator selection, and continuous optimization to refine the parameters associated with unary operators.

To address this mixed optimization problem, FEX employs a CO technique to determine the best selection of operators while leveraging a continuous optimization method to finetune the corresponding parameters of tree nodes. Regarding the CO, FEX utilizes a RL strategy to transform CO into a continuous optimization over probability distributions. To achieve this transformation, a controller network (modeled as a fully connected neural network) is integrated into FEX, which produces the probability mass function for each operator's selection at every tree node. Consequently, selecting the most suitable operators for tree nodes is recast as the problem of training the optimal controller network capable of sampling top-performing operators. The refinement process of identifying the best controller is fundamentally a continuous optimization task. The controller network's parameters are refined through policy gradient techniques to maximize the anticipated reward, consistent with RL conventions. As a result, the best operators can be efficiently identified by sampling from the probability distribution of the learned controller network.

In particular, analyzing dynamics in complex networks brings forth several challenges, primarily the computational complexity involved. Computing particle interactions on a graph requires $O(N^2)$ complexity,  where $N$ is the number of nodes in a network. This is  prohibitive for large-scale systems. We utilize a stochastic algorithm when calculating the loss function to mitigate this. Simply put, nodes in the network are grouped into random batches and particle interactions are limited to their respective batches. With this strategy, we achieve faster computational speeds, i.e., $O(N)$  computational complexity, without sacrificing the precision required to identify the accurate structure of the FEX tree.

\subsection{Functional space of finite expressions}
\label{sec:abstractframework}

FEX models the dynamics of network data within a functional space characterized by a finite set of operators. As such, it is crucial to clearly define this functional space where the solution resides.
\begin{definition}[Mathematical expression~\cite{liang2022finite}]\label{def:math_expression}\label{def:expression}
A mathematical expression is a combination of symbols, which is well-formed by syntax and rules and forms a valid function. The symbols include operands (variables and numbers), operators (e.g., \quotes{+}, \quotes{sin}, integral, derivative), brackets, and punctuation.
\end{definition}

\begin{definition}[$k$-finite expression~\cite{liang2022finite}]\label{def:k-finite}
A mathematical expression is called a $k$-finite expression if the number of operators in this expression is $k$.
\end{definition}

\begin{definition}[Finite expression method]\label{def:FEX}
The finite expression method aims to numerically capture dynamics by using a limited size expression, so that the produced function closely matches the actual dynamics.
\end{definition}

We denote $\mathbb{S}_k$ the functional space that consists of functions formed by finite expressions with the number of operators less than or equal to $k$.

\subsection{The combinatorial optimization problem in FEX}\label{sec:error} 
In FEX, the loss functional $\mathcal{L}$ varies based on the problem. For this study, we adopt the least squares loss functional on each dimension of the dynamics, as presented in~\cite{gao2022autonomous}, which is described as follows:

\begin{equation}
    \label{eqn:loss}
    \mathcal{L}(\theta_F, \theta_G) = \frac{1}{N T} \sum_{i=1}^{N} \sum_{t=1}^{T} \left\| \frac{d\mathbf{x}_i(t)}{dt} - \left( \mathbf{F}(\mathbf{x}_i(t); \theta_F) + \sum_{j=1}^{N} \mathbf{A}_{ij} \mathbf{G}(\mathbf{x}_i(t), \mathbf{x}_j(t); \theta_G) \right) \right\|_2^2,
\end{equation}
where $\mathbf{A}$ is the adjacency matrix of the network, $\mathbf{F}(\mathbf{x}_i(t); \theta_F)$ denotes the self-dynamics modeled by a FEX tree parameterized by $\theta_F$, $\mathbf{G}(\mathbf{x}_i(t), \mathbf{x}_j(t); \theta_G)$ denotes the interaction dynamics by another FEX tree parameterized by $\theta_G$. In FEX, the solution is found by solving the   combinatorial optimization problem 
\begin{align}
    \min_{\mathbf{F},\mathbf{G} \in \mathbb{S}_{\sf FEX}}\mathcal{L}(\mathbf{F},\mathbf{G}),
    \label{eqn:co}
\end{align}
where the solution space $\mathbb{S}_{\sf FEX}\subset\mathbb{S}_k$ will be elaborated in Section \ref{sec:binary_tree}.

\subsection{Implementation of FEX}
\label{sec:alg}

As mentioned previously, FEX's computational process begins with the formation of a mathematical binary tree, where each node represents either a unary or a binary operator. The candidate solution is derived from evaluating the function represented by this tree. Subsequently, the CO denoted by ~\eqref{eqn:co} is employed to dynamically choose the best operators for all tree nodes. This optimization aims to find operators that recover the structure of the genuine network dynamics.


\subsubsection{Finite expressions with binary trees}
\label{sec:binary_tree}
FEX uses a binary tree structure $\mathcal{T}$ to represent finite expressions. The user defines the sets of potential unary and binary operators, represented by $\mathbb{U}$ and $\mathbb{B}$. Common choices of unary and binary operators are respectively, 
$$
\sin,\exp, \text{Id}, (\cdot)^2, \int\cdot\text{d} x_i, \frac{\partial\cdot}{\partial x_i}, \cdots\
\quad{\rm and}\quad 
+,-,\times,\cdots.
$$
Every unary operator in the leaf level operates on the inputs in an element-wise manner, followed by scaling parameters $\alpha_i$, where $i=1,\ldots,d$, and a bias parameter $\beta$. For instance:
$$
\alpha_1\exp(x_1)+ \ldots +\alpha_d\exp(x_d) + \beta.
$$

\begin{figure}[ht]
    \centering
    \includegraphics[width=0.8\linewidth]{ 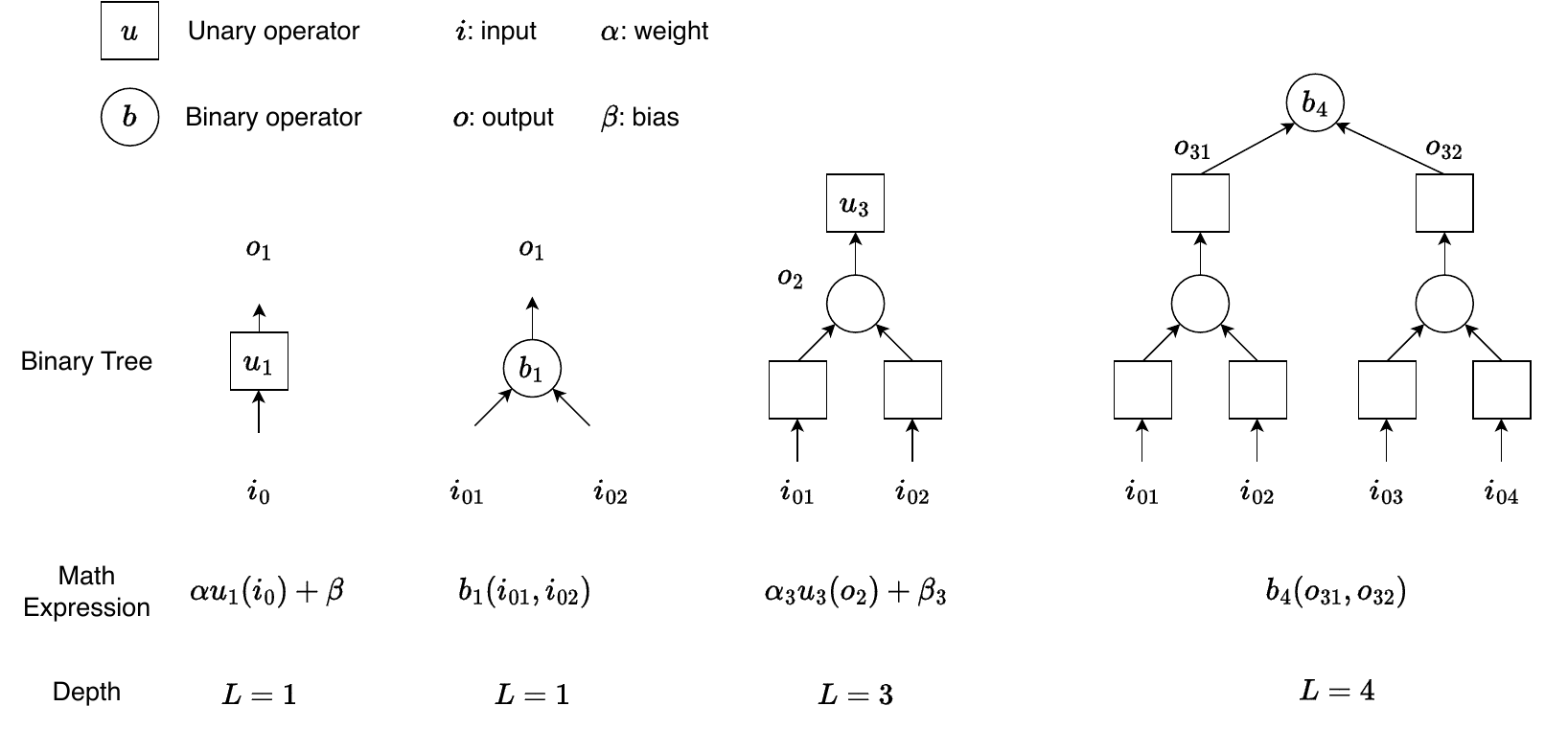}
    \caption{Computational rule of a FEX binary tree. Each unary operator is equipped with a weight $\alpha$ and bias $\beta$ to enhance expressiveness. For tree depth greater than one (i.e., $L > 1$), the computation is implemented by recursion.}

    \label{fig:fex_tree}
\end{figure}

All parameters associated with unary operators are collectively represented by $\bm{\theta}$. The complete expression is derived through a preorder traversal of the operator sequence, $\bm{e}$, in the binary tree $\mathcal{T}$. Consequently, this finite expression is represented as $u(\bm{x}; \mathcal{T}, \bm{e}, \bm{\theta})$, as a function of input $\bm{x}$. Given a fixed $\mathcal{T}$, the maximum number of operators remains constrained and is upper-bounded by a constant, labeled $k_{\mathcal{T}}$. In FEX, the functional space wherein the dynamics of the system is solved is defined as:
$$\mathbb{S}_{\sf FEX}=\{u(\bm{x}; \mathcal{T}, \bm{e}, \bm{\theta})~|~\bm{e}\in  \mathbb{U}\cup\mathbb{B}\} 
$$ 

The computation within the binary tree $\mathcal{T}$ proceeds recursively, beginning at the leaf nodes. Unary operators at these leaf nodes act on the input $\bm{x}$ in an element-wise manner, with the scaling factor $\alpha$, transforming the dimension from $\mathbb{R}^d$ to $\mathbb{R}$. Subsequently, computations proceed in a bottom-up recursive pattern until reaching the tree's root in the top level. A representation of the computation flow of such tree $\mathcal{T}$ is presented in Fig.~\ref{fig:fex_tree}.

\subsubsection{Workflow of FEX}
The FEX solution represented as 
$u(\bx;\mathcal{T},\bm{e},\bm{\theta})$ is derived by tackling the CO problem as:
\begin{align}
\min_{\bm{e}, \bm{\theta}} \mathcal{L}(u(\cdot; \mathcal{T}, \bm{e}, \bm{\theta})).
\label{eqn:obj}
\end{align}
To achieve this, FEX adopts a two-phase approach:

\textbf{Operator sequence optimization}: The primary goal here is to refine the operator sequence $\bm{e}$ such that it reflects the structure of the authentic dynamics.

\textbf{Parameter optimization}: Following the sequence refinement, FEX then optimizes the parameter set $\boldsymbol{\theta}$ aiming to minimize the functional described in~\eqref{eqn:obj}. The FEX framework is organized into four components:

\begin{itemize}
    \item \textit{Score Computation}: To discover the solution's structure, FEX deploys a mix-order optimization algorithm. This step assesses the score of the operator sequence $\bm{e}$. A stochastic algorithm is utilized to accelerate the score's calculation over the network data.
    \item \textit{Operator Sequence Generation}: Here, a neural network (NN) is leveraged to conceptualize a controller. This controller subsequently outputs a probability mass function, which is used for sampling optimal operator sequences.
    \item \textit{Controller Update}: This step refines the parameters of the controller based on the reward outcomes of the generated operator sequences, guiding it to yield more effective sequences over training iterations.
    \item \textit{Candidate Optimization}: FEX dynamically keeps a pool of the operator sequences with the highest rewards during the search process. Upon completion of training, each sequence candidate undergoes a fine-tuning phase. The ultimate objective is to identify the best operator sequence as the approximation of system dynamics.
\end{itemize}
Subsequent sections will elaborate on the details of each of these components.

\subsubsection{Score computation} \label{sec:score}
The evaluation of an operator sequence $\bm{e}$ through its score is pivotal during training. This score guides the controller's parameter updates, aiming to produce optimal probability mass functions for the sampling of efficient operators to approximate the network dynamics. The score of $\bm{e}$, denoted as $S(\Be)$, is defined as follows:
\begin{align}
    S(\Be) := \big(1+L(\Be)\big)^{-1}, \quad{\rm where}\quad L(\Be):=\min \{\mathcal{L}(u(\cdot; \mathcal{T}, \Be, \bm{\theta}))|\bm{\theta}\}.
    \label{eqn:orgscore}
\end{align}
For an efficient computation of the score $S(\Be)$, we employ a hybrid mix-order optimization strategy to update the parameter $\bm{\theta}$: We first take $T_1$ iterations using a first-order optimization algorithm, such as stochastic gradient descent~\cite{rumelhart1986learning} or Adam~\cite{kingma2014adam}, and this yields $\bm{\theta^{\Be}}_{T_1}$. $\bm{\theta^{\Be}}_{T_1}$ will be used as a warm start, followed by a  $T_2$ iterations of a second-order optimization method, such as Newton's method~\cite{avriel2003nonlinear} or BFGS~\cite{fletcher2013practical}, resulting in $\bm{\theta^{\Be}}_{T_1+T_2}$.

\textbf{Stochastic-FEX} To address the computational burden when performing the total $T_1 + T_2$ optimization steps, we propose to approximate the interaction terms using only a subset of the particles at each step. In particular, we divide the $N$ particles into $n$ batches and only consider the interactions within each batch, then the number of interactions that need to be computed is significantly reduced. Assuming that each batch contains approximately $p$ nodes ($n \approx \frac{N}{p}$), then the number of interactions to be computed for each batch is on the order of $p^2$. Since there are $n$ such batches, the total number of interactions at each step is on the order of $np^2 = \frac{N}{p} p^2 = Np$, which is $O(N)$ since $p$ is a constant independent of $N$. This significantly reduces the computational cost when evaluating the loss function during the $T_1 + T_2$ steps of optimization.

Upon completion of these steps, the score associated with the operator sequence $\Be$ is denoted as:
\begin{align}
    S(\Be) = \big(1+\mathcal{L} (u(\cdot; \mathcal{T}, \Be, \bm{\theta}_{T_1+T_2}^{\Be}))\big)^{-1}. 
    \label{eqn:score}
\end{align}
We have the following theorem on the convergence of Stochastic-FEX within this phase. The proof is provided in the Supplementary Information.
\begin{theorem}
    Fix some tolerance level \( \delta > 0 \), and integer $q \geq 2$. Let \(\theta^* = (\theta_F^*, \theta_G^*)\) be a regular minimizer of \(\mathcal{L}\) defined in \eqref{eqn:loss}, and suppose that Stochastic-FEX is run with a step-size schedule of the form \( \gamma_n = \gamma / (n + m)^p \) for some \( p \in (2/(q+2), 1] \) and large enough \( m, \gamma > 0 \). Furthermore, assume that the following bounds hold:
    
    \begin{enumerate}
        \item $\left\|\frac{d\mathbf{x}_i(t)}{dt} - \mathbf{F}(\mathbf{x}_i(t); \theta_F)\right\|_2 \leq C_1$
        \item $\left\|\mathbf{G}(\mathbf{x}_i(t), \mathbf{x}_j(t); \theta_G)\right\|_2 \leq M_1$
        \item $\left|\nabla_{\theta_G} \mathbf{G}(\mathbf{x}_i(t), \mathbf{x}_j(t); \theta_G)_k\right| \leq M_2$, where $(\cdot)_k$ denotes the $k$-th component of the vector.
    \end{enumerate}
    
    Then:
    \begin{enumerate}
        \item There exist neighborhoods \(\mathcal{U}\) and \(\mathcal{U}_1\) of \(\theta^*\) such that, if \(\theta_1 \in \mathcal{U}_1\), the event
        \[
        \Omega_{\mathcal{U}} = \{\theta_n \in \mathcal{U} \text{ for all } n = 1, 2, \ldots\}
        \]
        occurs with probability at least \( 1 - \delta \).
        
        \item Conditioned on \(\Omega_{\mathcal{U}}\), we have
        \[
        \mathbb{E}[\|\theta_n - \theta^*\|_2^2 \mid \Omega_{\mathcal{U}}] = O(1/n^p).
        \]
    \end{enumerate}
\end{theorem}

\subsubsection{Operator sequence generation}
The primary objective of the controller is to generate operator sequences that achieve high scores throughout the training process. We denote the controller as $\bm{\chi}_\Phi$, which is parameterized by a fully connected neural network with parameters $\Phi$. Given an operator sequence $\Be$ comprising $s$ nodes, the controller $\bm{\chi}_\Phi$ produces probability mass functions $\bm{p_{\Phi}}^i$ for $i = 1, \ldots, s$. The operator $e_j$ is then sampled based on the probability mass function $\bm{p_{\Phi}}^j$. To encourage exploration within the set of operators, the $\epsilon$-greedy strategy~\cite{sutton2018reinforcement} is implemented. To be more specific, the operator $e_i$ is selected from a uniform distribution over the entire operator set with a probability of $\epsilon<1$.

\subsubsection{Controller update}
As previously discussed, we introduce a controller represented by the neural network $\bm{\chi}_\Phi$, to generate a probability mass function for each node in the binary tree. This transformation changes the problem from a CO problem into a continuous optimization one. The primary goal now becomes the optimization of the controller parameters $\Phi$ so that the controller $\bm{\chi}_\Phi$ is capable of producing optimal probability mass function for each node. To achieve this, we consider the expected value of scores of operator sequences sampled from the controller $\bm{\chi}_\Phi$, i.e., 
\begin{align}
    \mathcal{J}(\Phi):=\mathbb{E}_{\Be \sim \bm{\chi}_\Phi} S(\Be).
    \label{eqn:expect}
\end{align}
The derivative of \eqref{eqn:expect} with respect to $\Phi$ is 
\begin{align}
    \nabla_\Phi\mathcal{J}(\Phi)=\mathbb{E}_{\Be \sim \bm{\chi}_\Phi} \left\{S(\Be)\sum_{i=1}^s \nabla_\Phi \log\left(\bm{p}_\Phi^i(e_i)\right)\right\},
    \label{eqn:expectgrad}
\end{align}
where $\Bp_\Phi^i(e_i)$ is the probability of the sampled $e_i$. To efficiently approximate the expectation~\eqref{eqn:expectgrad}, we sample a batch of operator sequences $\{\Be^{(1)}, \Be^{(2)}, \cdots, \Be^{(M)}\}$ each time and compute
\begin{align}
    \nabla_\Phi\mathcal{J}(\Phi)\approx \frac{1}{M}\sum_{k=1}^M \left\{S(\Be^{(k)})\sum_{i=1}^s \nabla_\Phi \log\left(\Bp_\Phi^i(e_i^{(k)})\right)\right\},
    \label{eqn:mcmcavg}
\end{align}
where $M$ is the batch size. Subsequently, the model parameter $\Phi$ is updated using gradient ascent, denoted as $\Phi \leftarrow \Phi+\eta \nabla_\Phi\mathcal{J}(\Phi)$. Nonetheless, the practical aim is to identify the operator sequence $\Be$ with the highest score, rather than optimizing the average scores of all generated operator sequences. Therefore, we follow the approach in~\cite{petersen2021deep} and consider
\begin{align}
    \mathcal{J}(\Phi)=\mathbb{E}_{\Be \sim \bm{\chi}_\Phi} \left\{S(\Be)|S(\Be)\geq S_{\nu, \Phi}\right\},
    \label{eqn:expectriskseeking}
\end{align}
where $S_{\nu, \Phi}$ represents the $(1-\nu)\times 100\%$-quantile of the score distribution generated by $\bm{\chi}_{\Phi}$. In a discrete form,
the gradient computation becomes
\begin{align}
    \nabla_\Phi\mathcal{J}(\Phi)\approx \frac{1}{M}\sum_{k=1}^M \left\{(S(\Be^{(k)})-\hat{S}_{\nu, \Phi})\mathbf{1}_{\{S(\Be^{(k)})\geq \hat{S}_{\nu, \Phi}\}}\sum_{i=1}^s \nabla_\Phi \log\left(\Bp_\Phi^i(e_i^{(k)})\right)\right\},
    \label{eqn:mcmcrisk}
\end{align}
where $\mathbf{1}$ denotes an indicator function which evaluates to $1$ when the condition is met, and $0$ otherwise. Meanwhile, $\hat{S}_{\nu, \Phi}$ represents the $(1-\nu)$-quantile of the scores within the set $\{S(\mathbf{e}^{(i)})\}_{i=1}^M$.

\subsubsection{Candidate optimization}
During the training phase of optimizing the controller parameters $\Phi$, efficient evaluation of expressions sampled from the controller is crucial. As introduced in Section~\ref{sec:score}, this is achieved by computing the score $S(\mathbf{e})$ through $T_1 + T_2$ ``coarse-tune'' iterations. However, this coarse score might not reflect the performance of operator sequence $\Be$ owing to the optimization of a nonconvex function. Therefore, we maintain a pool $\mathbb{P}$ with a fixed size of $K$ during the training of controller, which dynamically holds the top-performing candidate expressions. When the controller's training is completed, we further refine the coefficients of each candidate expression $\Be$ within the pool $\mathbb{P}$ to attain high accuracy. To enhance the robustness of our method against link perturbations, we employ a random sampling strategy. In particular, for each candidate function, we perform random sampling of $L$ times, where we randomly sample $S$ nodes from the graph each time. For each batch, we perform $T_3$ iterations of optimizations based on the objective function $\mathcal{L}(u(\cdot; \mathcal{T}, \Be, \bm{\theta}))$ with a first-order algorithm with a small learning rate, and the final coefficients $\hat{\bm{\theta}}$ of the expression are obtained by averaging the coefficients over $L$ times.

\section{Dynamics learning results}\label{sec:results}
In this section, we present the results of identification of HR, FHN, and coupled R\"ossler dynamics with FEX on the Scale-Free (SF) network with clean data. The set up of SF network follows the description in Supplementary Information. We choose SF network since it demonstrates more complex phenomena than the ER model, and many real-world networks have scale-free property. During the offline stage, we generate the time series data with the true dynamics (ODEs). During the inference stage, we use FEX only with the time series data to identify the governing equations of the system, without accessing to the genuine ODEs. We emphasize that we only have access to the time series data and approximate the state time derivatives using numerical derivatives. This contrasts with the SINDy method, which requires direct access to the time derivatives of the states. Throughout all numerical experiments, we set the batch size in stochastic-FEX as 32. In addition, we provide the depths of FEX trees used in each of the considered dynamics in the numerical examples in Table~\ref{tab:tree_depth}. Robustness analyses of FEX and other baseline methods, aimed at extracting dynamics from low-resolution data, networks with spurious or missing links, and noisy data, are discussed in the Supplementary Information.
\begin{table}
    \centering
    \begin{tabular}{cc}
        \toprule
        Dynamics & Depth of Tree \\
        \midrule
         HR & 4  \\
         FHN & 3  \\
         Coupled R\"ossler & 3 \\
        \bottomrule
    \end{tabular}
    \caption{Tree depth of FEX tree for each dynamics.} 
    \label{tab:tree_depth}
\end{table}

\subsection{Numerical derivatives and error metric}
To approximate the time-varying derivative of each node, we apply the five-point approximation~\cite{sauer2011numerical}
\begin{equation}
\label{eqn:derivative}
\dot{x}_t \approx \frac{x_{t-2 \delta t}-8 x_{t-\delta t}+8 x_{t+\delta t}-x_{t+2 \delta t}}{12 \delta t},
\end{equation}
where $\delta t$ is the time step.

To evaluate the accuracy of inferred dynamics, we use the metric of symmetric mean absolute percentage error (sMAPE)~\cite{flores1986pragmatic}:

\begin{equation}
    \label{eqn:sMAPE}
    \mathrm{sMAPE}=\frac{1}{m} \sum_{i=1}^m \frac{\left|D_i-R_i\right|}{\left(\left|D_i\right|+\left|R_i\right|\right)},
\end{equation}
where $m$ is the total number of terms of both true and inferred functions, $D_i$ is the inferred coefficient and $R_i$ is the true coefficient. sMAPE is an appropriate metric in evaluating the performance of baselines since it not only captures the coefficient discrepancies but also penalizes incorrect inferred terms. sMAPE ranges from 0 to 1, and the lower the value the higher the accuracy, and vice versa.

\subsection{Hindmarsh-Rose dynamics}
The HR model~\cite{hindmarsh1982model,storace2008hindmarsh,wang1993genesis} was introduced by Hindmarsh and Rose in the early 1980s to describe neuronal activity. The HR model is a simplified model of Hodgkin-Huxley model~\cite{guckenheimer2002chaos} while still reproducing many of the dynamic behaviors observed in real neurons. The governing equations of HR model is given as

\begin{equation}
    \label{eqn:HR}
    \left\{\begin{array}{l}
\frac{d x_{i, 1}}{d t}=x_{i, 2}-a x_{i, 1}^3+b x_{i, 1}^2-x_{i, 3}+I_{e x t}+\epsilon\left(V_{\mathrm{syn}}-x_{i, 1}\right) \sum_{j=1}^N \mathbf{A}_{i j} \sigma\left(x_{j, 1}\right), \\
\frac{d x_{i, 2}}{d t}=c-u x_{i, 1}^2-x_{i, 2}, \\
\frac{d x_{i, 3}}{d t}=r\left[s\left(x_{i, 1}-x_0\right)-x_{i, 3}\right],
\end{array}\right.
\end{equation}
where the coupling term is
\begin{equation*}
    \sigma\left(x_{j, 1}\right)=\frac{1}{1+e^{-x_{j, 1}}} .
\end{equation*}
Here $x_{i,1}$ is the membrane potential of neuron $i$, $x_{i,2}$ is the transport rate of ions across the membrane through the ion channels, and $x_{i,3}$ is the adaptation current. Parameters are set as: $a = 1, b = 3, c = 1, u = 5, s = 4,r = 0.004, x_0 = -1.6$, $\epsilon = 0.15, V_{syn} = 2$, and $I_{ext} = 3.24$ is external current. Notice that the dynamic
behavior of the HR neuron can be adjusted by varying these parameters. By doing so, one can observe various neuronal activities, such as tonic spiking, phasic spiking and chaotic bursting. We generate the time series data with Equation~\eqref{eqn:HR}, with the adjacency matrix of SF network, terminal time $T = 500$ and time step $\delta t = 0.01$.

Applying FEX to the neuronal activities data generated by HR dynamics on a directed SF network, we obtain the inferred equations as

\begin{equation}
\label{eqn:HR_FEX}
    \left\{\begin{aligned}
\frac{d \hat{x}_{i, 1}}{d t} & =0.9961 x_{i, 2}-1.0014 x_{i, 1}^3+2.9758 x_{i, 1}^2-0.9917 x_{i, 3}+3.2392 \\
& +\sum_{j=1}^N \mathbf{A}_{i j}\left(0.2985  \sigma \left(x_{j, 1}\right)-0.1477x_{i,1}\sigma \left(x_{j, 1}\right)\right), \\
\frac{d \hat{x}_{i, 2}}{d t} & =1.0004-5.0001 x_{i, 1}^2-1.0001 x_{i, 2}, \\
\frac{d \hat{x}_{i, 3}}{d t} & =0.0255+0.0182 x_{i, 1}-0.0050 x_{i, 3},
\end{aligned}\right.
\end{equation}

\begin{figure}[!ht]
    \centering
    \includegraphics[width=1\linewidth]{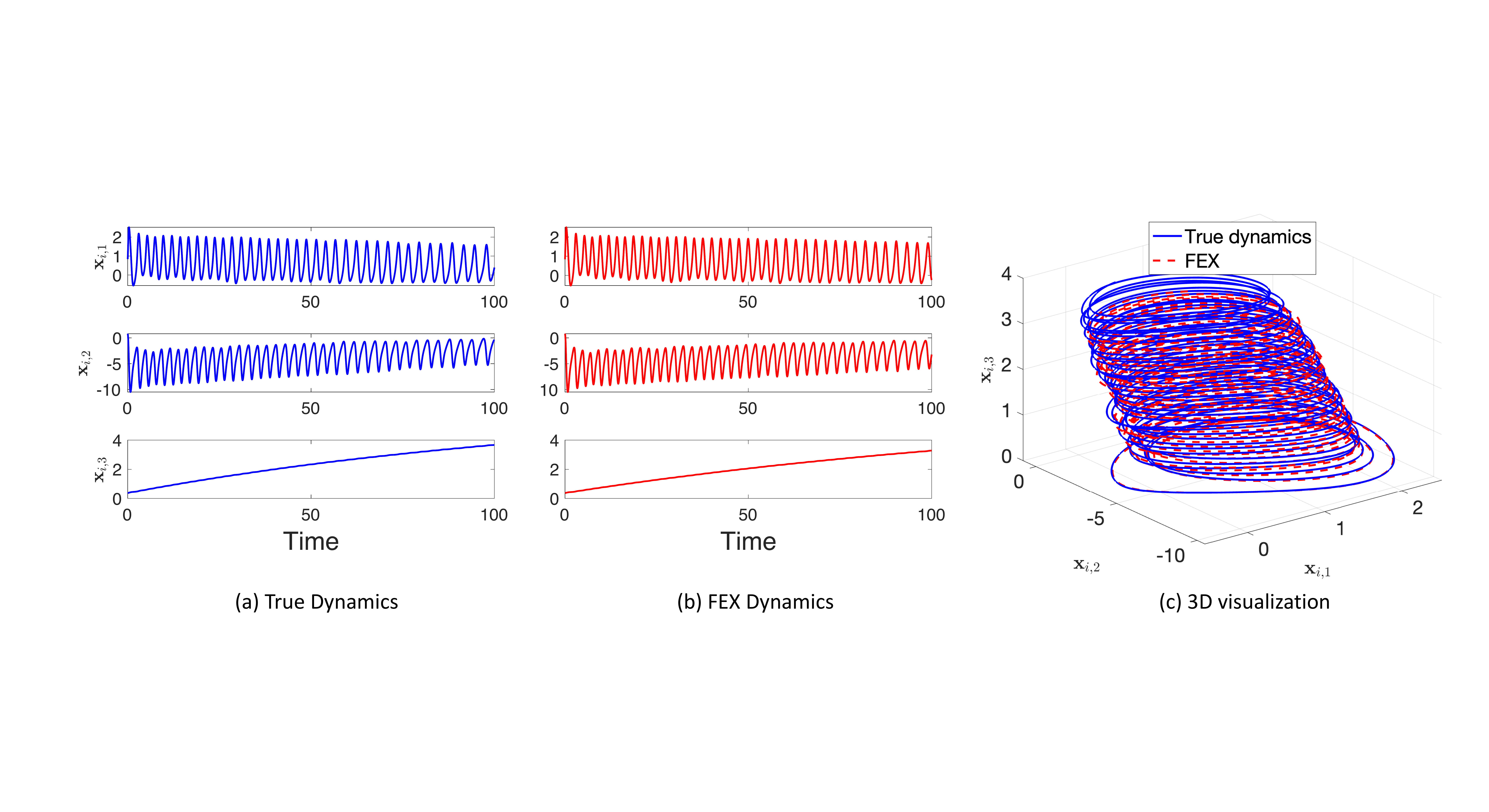}
    \caption{True HR dynamics and the dynamics identified by FEX.}

    \label{fig:hr}
\end{figure}

From Eq.~\eqref{eqn:HR_FEX}, we observe that FEX not only identifies all the correct terms in the dynamics in each dimension of the node, but also learns the coefficient of each term with relative errors lower than $1\%$. The neuronal activities generated by FEX and true governing equations are shown in Fig.~\ref{fig:hr}.

\subsubsection{Observational noises}
In this subsection, we demonstrate the robustness of our method against observational noises. We focus on the HR dynamics on a directed SF network. We generate noisy data through
\begin{equation*}
    X_i^{o b s}(t)=\boldsymbol{x}_i(t)+a \beta_X(t),
\end{equation*}
where $\beta_X(t)$ is observational noise, which follows a Gaussian distribution with zero mean and standard deviation one, so the parameter $a$ is used to tune the level of noise. 


\begin{table}
    \centering
    \begin{tabular}{lllll}
        \toprule
        & True Dynamics &  Two-Phase &  ARNI & FEX \\
        \midrule
        $\mathbf{F}(x_i)$ &$ x_{i,2} $ & $0.9631 x_{i,2}$ & $0.6002x_{i,2}$ & $0.9916 x_{i,2}$ \\
        \hline
        &$-x_{i,1}^3$ & $-1.0039x_{i,1}^3$ & $-1.3202x_{i,1}^3$&$-1.0023x_{i,1}^3$   \\
        \hline
        &$3x_{i,1}^2$ &$2.7820x_{i,1}^2$  &$2.3557x_{i,1}^2$ & $2.9659x_{i,1}^2$  \\
        \hline
        &$-x_{i,3}$ &$-0.9663x_{i,3}$  & $-0.7982x_{i,3}$& $-0.9871x_{i,3}$  \\
        \hline
        &$3.24$ &$2.9673$  &$1.9699$ & $3.2453$  \\
        \hline
        && &$0.6597\exp(x_{i,1})$ &  \\
        \hline
        $\mathbf{G}(x_i,x_j)$ &$0.3\sigma(x_{j,1})$ & $0.1529\sigma(x_{j,1})$ &$0.2804\sigma(x_{j,1})$ & $0.2963\sigma(x_{j,1})$\\
        \hline
       &$-0.15 x_{i,1}\sigma(x_{j,1})$ &$-0.0995 x_{i,1}\sigma(x_{j,1})$  & $-0.0847x_{i,1}\sigma(x_{j,1})$ & $-0.1262 x_{i,1}\sigma(x_{j,1})$\\

        \bottomrule
    \end{tabular}
    \caption{Numerical results for Hindmarsh-Rose dynamics (SNR = 45) in self-dynamics $\bF(x_i)$ and interaction dynamics $\bG(x_i, x_j)$ term by term.}.
    \label{tab:hr_snr_45}
\end{table}

\begin{table}
    \centering
    \begin{tabular}{lllll}
        \toprule
        & True Dynamics &  Two-Phase &  ARNI & FEX \\
        \midrule
        $\mathbf{F}(x_i)$ &$ x_{i,2} $ & $ 0.9341x_{i,2}$ & $0.3916x_{i,2}$ & $0.9633x_{i,2}$ \\
        \hline
        &$-x_{i,1}^3$ & $-0.9948x_{i,1}^3$ & $-1.0638x_{i,1}^3$&$-1.0058x_{i,1}^3$   \\
        \hline
        &$3x_{i,1}^2$ &$2.6599x_{i,1}^2$  &$2.5882x_{i,1}^2$ & $2.8923x_{i,1}^2$  \\
        \hline
        &$-x_{i,3}$ &$-0.9332x_{i,3}$  & $-0.3871x_{i,3}$& $-0.9588x_{i,3}$  \\
        \hline
        &$3.24$ &$3.1341$  &$1.8715$ & $3.2342$  \\
        \hline
        && &$0.1318x_{i,2}x_{i,3}$ &  \\
        \hline
        && &$0.2246\exp(x_{i,2})$ &  \\
        \hline
        $\mathbf{G}(x_i,x_j)$ &$0.3\sigma(x_{j,1})$ & $0.1333\sigma(x_{j,1})$ &None & $0.2683\sigma(x_{j,1})$\\
        \hline
       &$-0.15x_{i,1}\sigma(x_{j,1})$ & None & None & $-0.1118x_{i,1}\sigma(x_{j,1})$\\
        \hline
        & &  & $-0.2341\sigma(x_{j,1}-x_{i,1})$ & \\
        \bottomrule
    \end{tabular}
    \caption{Numerical results for Hindmarsh-Rose dynamics (SNR = 40) in self-dynamics $\bF(x_i)$ and interaction dynamics $\bG(x_i, x_j)$ term by term.}.
    \label{tab:hr_snr_40}
\end{table}

\begin{table}
    \centering
    \begin{tabular}{clll}
        \toprule
        SNR & Two-Phase &  ARNI & FEX \\
        \midrule
         $45$ & 0.0924 & 0.2720 & 0.0158 \\
         \hline
         $40$ & 0.2190 & 0.6252 & 0.0376 \\
        \bottomrule
    \end{tabular}
    \caption{sMAPE of HR dynamics with different levels of noise.} 
    \label{tab:hr_mse}
\end{table}

From Table~\ref{tab:hr_snr_45}, \ref{tab:hr_snr_40}  and \ref{tab:hr_mse}, FEX demonstrates its capability to accurately identify all terms within the HR dynamics with their respective coefficients with SNR = 45 and 40. The Two-Phase method can indeed infer the correct terms in the dynamics in the case of SNR = 45. However, the accuracy of inferred coefficients is low, particularly in the interaction dynamics. In addition, in the case of SNR = 40, Two-Phase method fails to infer the $-0.15x_{i,1}\sigma(x_{j,1})$ term in the interaction dynamics. ARNI can infer most of the terms correctly when SNR = 45, with an extra term $0.6597\exp(x_{i,1})$ in the self-dynamics. Furthermore, the coefficients in the self-dynamics are poorly estimated, but it is interesting to notice that the coefficients of interaction-dynamics are more accurate than those of Two-Phase, owing again to ARNI's capability to infer the network structure. As for SNR = 40, ARNI infers more redundant terms in self-dynamics, and it fails to identify both true terms in the original dynamics, and incorrectly identifies a $-0.2341\sigma(x_{j,1} - x_{i,1})$ term. This shows that ARNI can reasonably well infer most of the dynamics terms in low-noise case, while it fails to work as noise level increases. Therefore, FEX has demonstrated its superior performance compared to Two-Phase and ARNI to handle noisy data.

\subsection{FitzHugh-Nagumo dynamics}
We further test our method to the neuronal activities data generated by FHN dynamics~\cite{fitzhugh1961impulses}. The FHN model is another mathematical representation used to describe the spiking behavior of neurons. It is defined by a system of two ordinary differential equations, capturing the
main features of excitability in nerve membrane dynamics. The equations governing the FHN neuronal network dynamics are
\begin{equation}
    \label{eqn:FHN}
    \left\{\begin{array}{l}
\frac{d x_{i, 1}}{d t}=x_{i, 1}-x_{i, 1}^3-x_{i, 2}-\epsilon \sum_{j=1}^N \mathbf{A}_{i j} \frac{\left(x_{j, 1}-x_{i, 1}\right)}{k_i^{i n}}, \\
\frac{d x_{i, 2}}{d t}=a+b x_{i, 1}+c x_{i, 2},
\end{array}\right.
\end{equation}
where the first component $x_{i,1}$ represents the membrane potential containing self and interaction dynamics, $k_i^{i n}$ is the in-degree of neuron $i$ (representing the number of incoming connections to node $i$), and $\epsilon = 1$. The second component $x_{i,2}$ represents a recovery variable where $a = 0.28, b = 0.5$ and $c = -0.04$. We generate the time series data with Equation~\eqref{eqn:FHN}, with the adjacency matrix of SF network, terminal time $T = 300$ and time step $\delta t = 0.01$.

The equations inferred by our approach from the neuronal activities data generated on a directed SF network are shown below:
\begin{equation*}
\label{eqn:fnd_fex}
    \left\{\begin{array}{l}
\frac{d \hat{x}_{i, 1}}{d t}=0.9942 x_{i, 1}-0.9998 x_{i, 1}^3-0.9999 x_{i, 2}-1.0022 \sum_{j=1}^N A_{i j} \frac{\left(x_{j, 1}-x_{i, 1}\right)}{k_i^{\text {in }}} \\
\frac{d \hat{x}_{i, 2}}{d t}=0.2801+0.5000 x_{i, 1}-0.0400 x_{i, 2} .
\end{array}\right.
\end{equation*}

\begin{figure}[!ht]
    \centering
    \includegraphics[width=1.0\linewidth]{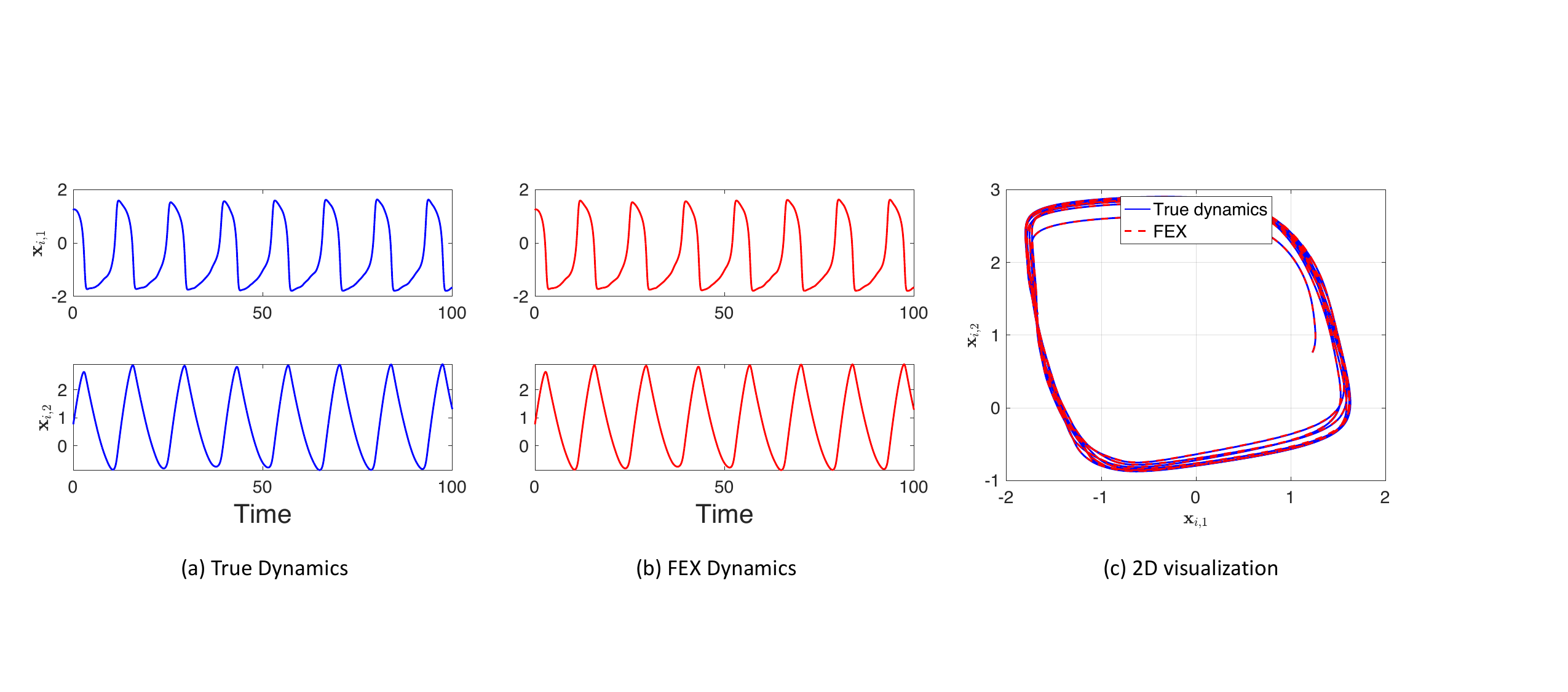}
    \caption{True FHN dynamics and the dynamics identified by FEX.}

    \label{fig:fhn}
\end{figure}

The trajectories generated by physical laws inferred by FEX and the true governing equations are shown in Fig.~\ref{fig:fhn}. FEX has shown to accurately infer both dimensions of FHN system and coincide closely with the true dynamics.

We compare the robustness of FEX against spurious and missing links using FHN dynamics, and evaluate it alongside Two-Phase and ARNI methods. FEX demonstrates superior accuracy compared to the baselines, even with random addition or deletion of $15\%$ of the links. Detailed results are provided in the Supplementary Information.


\subsection{Coupled R\"ossler dynamics}
We also consider the coupled R\"ossler oscillators~\cite{rosenblum1996phase,tang2019master}, which is a classical model often used to study chaotic dynamics and synchronization phenomena in complex networks. We generate chaotic activities data according to the true equations governing heterogeneous R\"ossler dynamics~\cite{barzel2015constructing,schmidt2009distilling},
\begin{equation}
\label{eqn:ross}
    \left\{\begin{array}{l}
\frac{d x_{i, 1}}{d t}=-\omega_i x_{i, 2}-x_{i, 3}+\epsilon \sum_{j=1}^n A_{i j}\left(x_{j, 1}-x_{i, 1}\right), \\
\frac{d x_{i, 2}}{d t}=\omega_i x_{i, 1}+a x_{i, 2}, \\
\frac{d x_{i, 3}}{d t}=b+x_{i, 3}\left(x_{i, 1}+c\right),
\end{array}\right.
\end{equation}
where coupling strength $\epsilon = 0.15$, $a = 0.2, b = 0.2,$ and $c = -5.7$. The natural frequencies of the oscillators, denoted by $\omega_i$, follow a normal distribution with a mean value of 1 and a standard deviation of 0.1. We generate the time series data with Equation~\eqref{eqn:ross}, with the adjacency matrix of SF network, terminal time $T = 100$ and time step $\delta t = 0.01$.

\begin{figure}[!ht]
    \centering
    \includegraphics[width=1.0\linewidth]{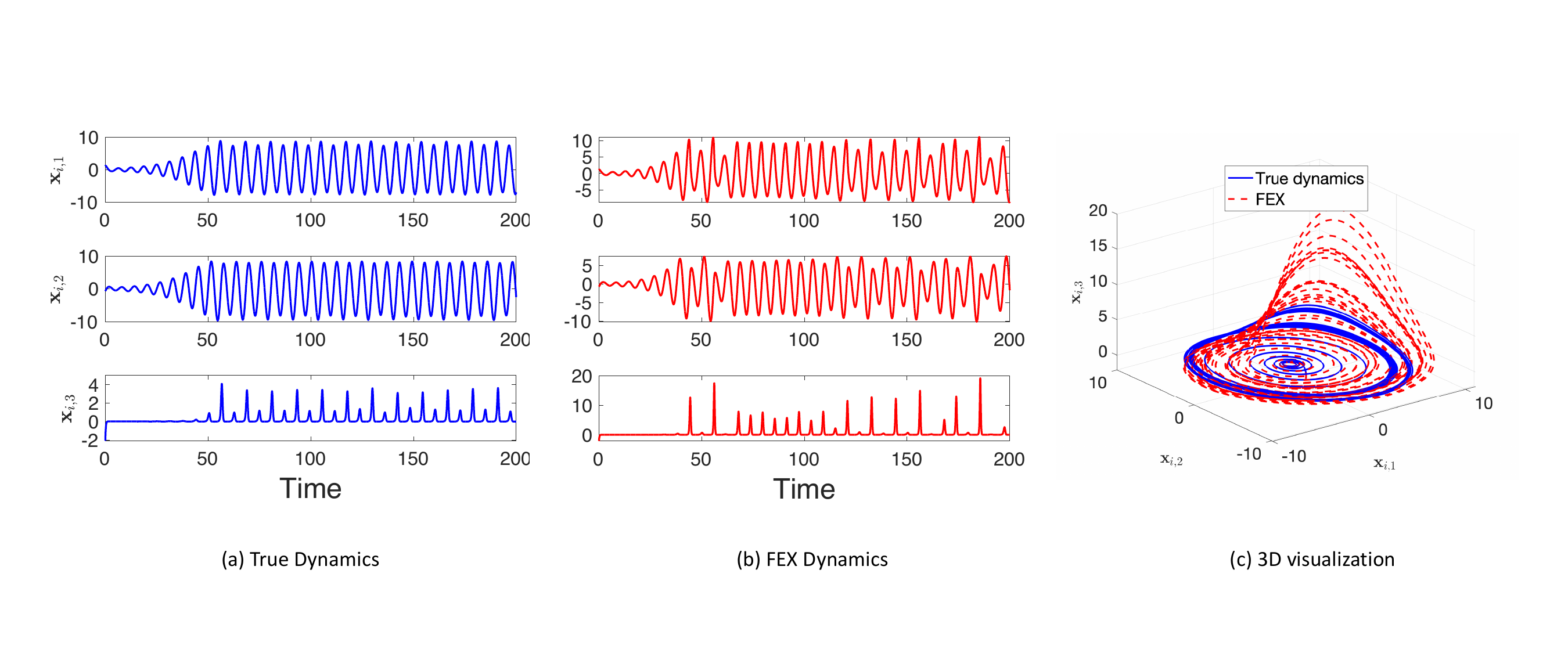}
    \caption{True coupled R\"ossler dynamics and the dynamics identified by FEX.}

    \label{fig:ross}
\end{figure}

The equations inferred by FEX from the data generated on a directed SF network with $\epsilon = 0.15$ are
\begin{equation}
    \label{eqn:ross_fex}
    \left\{\begin{array}{l}
\frac{d \hat{x}_{i, 1}}{d t}=-1.0093 x_{i, 2} - 1.0027 x_{i, 3}+0.1491\sum_{j=1}^n A_{i j}\left(x_{j, 1}-x_{i, 1}\right), \\
\frac{d \hat{x}_{i, 2}}{d t}=0.9909 x_{i, 1}+0.2030 x_{i, 2}, \\
\frac{d \hat{x}_{i, 3}}{d t}=0.1967+0.9987 x_{i, 3} x_{i, 1}-5.6653 x_{i, 3},
\end{array}\right.
\end{equation}
Note that the coefficient -1.0093 of $x_{i,2}$ in the first equation of Eq.~\eqref{eqn:ross_fex} and 0.9909 of $x_{i,1}$ in the second equation of Eq.~\eqref{eqn:ross_fex} are estimated average value of nodes' natural frequencies, otherwise inferring a distinctive frequency for each node would lead to an $n-$fold increase in the dimensionality of model space. The trajectories generated by physical laws inferred by FEX and the true governing equations are shown in Fig.~\ref{fig:ross}.

Utilizing Coupled R\"ossler dynamics, we further assess the robustness of FEX against low-resolution observational data. We compare the performance of FEX with Two-Phase and ARNI methods when observing only $10\%$ and $5\%$ of the time series data. FEX demonstrates superior accuracy in identifying the dynamics within this low data regime compared to the other baselines. Detailed results are provided in the Supplementary Information.

\subsection{Details of performance of stochastic-FEX}

As introduced before, computation on a large complex network is demanding or even infeasible. Therefore, we incorporate a stochastic algorithm into the FEX algorithm to accelerate the identification of the optimal mathematical structure of network dynamics while maintaining a high accuracy.

\begin{figure}[!ht]
    \centering
    \includegraphics[width=1\linewidth]{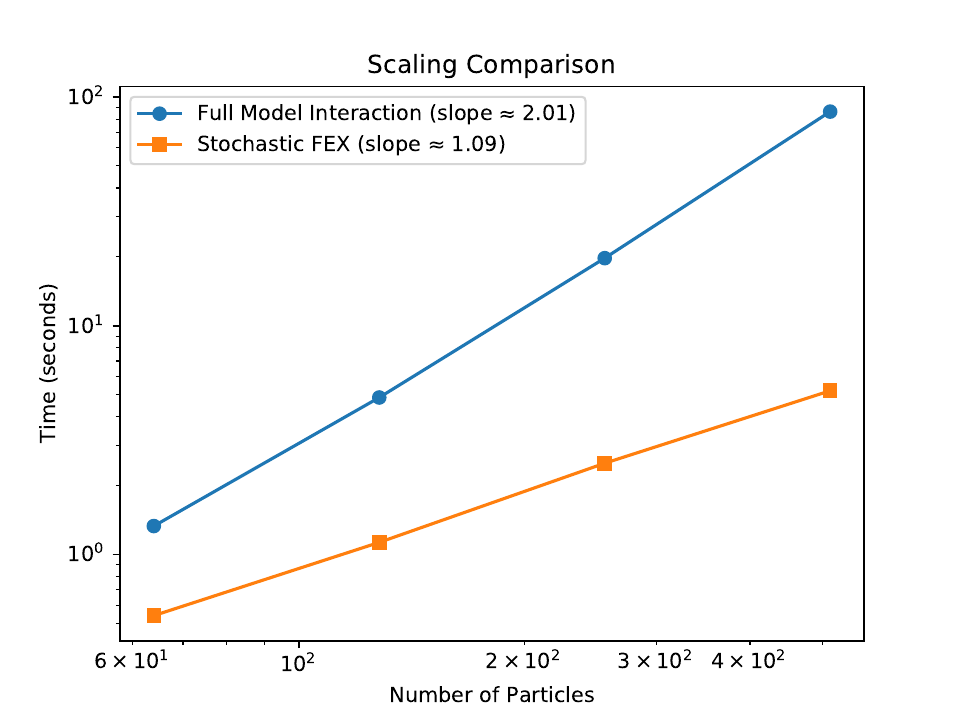}
    \caption{Clock time comparisons of full interaction model and stochastic-FEX. The $x$-axis shows 64, 128, 256, and 512 particles of the SF network, respectively. The $y$-axis shows the clock time of one iteration computation for each network size.}

    \label{fig:rbm}
\end{figure}

To demonstrate the efficiency, we create a SF model of network size $64, 128, 256$, and $512$, respectively. Then, we use the time series data of FHN model on the four SF networks, and implement FEX to infer the dynamics. We record the clock time of computation of $T_1 + T_2$ coarse-tune in the FEX training algorithm by full interaction method and stochastic-FEX (with a batch size of 32), respectively. The log-log plot of time vs. number of particles is shown in Fig.~\ref{fig:rbm}. We observe that the fitting line clock time of the full interaction model on different network sizes has a slope of approximately 2. The stochastic-FEX's slope is roughly 1, indicating that the stochastic-FEX demonstrates a linear scaling performance. 

Furthermore, we comment that we only use this stochastic algorithm in the CO part of FEX to identify the optimal structure of the dynamics. The CO stage is the computational bottleneck of FEX, and we have demonstrated its capability to determine the correct structure of dynamics while accelerating the computation significantly. After the structures of top-performing candidates are learned, we use the full interaction model in the $T_3$ iterations of coefficients fine-tuning stage to achieve high accuracy of dynamics inferred by FEX. The computational burden at this stage is manageable since we only have to fine-tune the coefficients of the top $K$ candidates in the pool $\mathbb{P}$.

\section{Conclusion}\label{sec:conclusion}
 In this work, we proposed FEX and its fast algorithm as a novel methodology to learn the dynamics on complex networks. FEX has shown its capability to accurately discover physical laws on various synthetic networks, even with noise and low resolution of data. Furthermore, we incorporate a stochastic algorithm into our proposed FEX framework, which scales up FEX to handle large data when the network size increases significantly. Therefore, we have demonstrated that FEX has the potential to identify real-world dynamics in an interpretable and reliable way. 

 There are also questions that remained to be addressed. First, throughout our work, we consider only the pairwise interaction dynamics in the physical laws. However, it is common for complex systems to exhibit three-point or higher-order couplings among nodes in the network. Yet, it is straightforward to adapt such a setting into the FEX methodology, with the extra computational cost of including more binary trees to represent higher-order dynamics. Second, dynamics often demonstrate implicit stochasticity in real-world data, which can be described by stochastic differential equations (SDEs)~\cite{deco2009stochastic,genkin2021learning}. Lastly, this work focuses on classical dynamics with an explicit form of governing equations. However, dynamics are largely unknown or underexplored in most situations. Hence, it is interesting to observe how accurately FEX can approximate empirical dynamics and the insights it provides from inferring complex network dynamics.

 \textbf{Acknowledgments}
 C. W. was partially supported by National Science Foundation under award  DMS-2206332. H. Y. was partially supported by the US National Science Foundation under awards DMS-2244988, DMS-2206333, the Office of Naval Research Award N00014-23-1-2007, and DARPA D24AP00325-00.

 \textbf{Competing interests}
 The authors declare no competing interests.







\bibliography{main}
\bibliographystyle{plain}




\end{document}


\title{\textbf{Supplementary Information} for: \\ A Fast Algorithm for the Finite Expression Method in Learning Dynamics on Complex Networks}

\author[1]{Zezheng Song\thanks{Email: \texttt{zsong001@umd.edu}.}}
\author[2]{Chunmei Wang\thanks{Email: \texttt{chunmei.wang@ufl.edu}.}}
\author[1,3]{Haizhao Yang\thanks{Email: \texttt{hzyang@umd.edu}.} }

\affil[1]{Department of Mathematics, University of Maryland, College Park, MD 20742, USA}
\affil[2]{Department of Mathematics, University of Florida, Gainesville, FL 32611, USA   }
\affil[3]{Department of Computer Science, University of Maryland, College Park, MD 20742, USA}

\date{}
\maketitle







\maketitle
\tableofcontents

\section{Convergence of Stochastic-FEX algorithm}

In this section, we provide proof for Theorem 1 in the main text. Recall we consider the following optimization problem:
\begin{equation*}
    \mathcal{L}(\theta_F, \theta_G) = \frac{1}{N T} \sum_{i=1}^{N} \sum_{t=1}^{T} \left | \frac{d\mathbf{x}_i(t)}{dt} - \left( \mathbf{F}(\mathbf{x}_i(t); \theta_F) + \sum_{j=1}^{N} \mathbf{A}_{ij} \mathbf{G}(\mathbf{x}_i(t), \mathbf{x}_j(t); \theta_G) \right) \right |_2^2,
\end{equation*}
where $\mathbf{A}$ is the adjacency matrix of the network, $\mathbf{F}(\mathbf{x}_i(t); \theta_F)$ denotes the self-dynamics modeled by a FEX tree parameterized by $\theta_F$, $\mathbf{G}(\mathbf{x}_i(t), \mathbf{x}_j(t); \theta_G)$ denotes the interaction dynamics by another FEX tree parameterized by $\theta_G$. We introduce the following notations.
\[
\begin{aligned}
    & a_i = \frac{d \mathbf{x}_i(t)}{dt} - \mathbf{F}(\mathbf{x}_i(t); \theta_F), \\
    & b_i^{\text{full}} = \sum_{j=1}^{N} \mathbf{A}_{ij} \mathbf{G}(\mathbf{x}_i(t), \mathbf{x}_j(t); \theta_G), \\
    & b_i^{\text{mini}} = \frac{N}{S} \sum_{j \in B_1} \mathbf{A}_{ij} \mathbf{G}(\mathbf{x}_i(t), \mathbf{x}_j(t); \theta_G), \\
    & \nabla_{\theta_G} b_i^{\text{full}} = \sum_{j=1}^{N} \mathbf{A}_{ij} \nabla_{\theta_G} \mathbf{G}(\mathbf{x}_i(t), \mathbf{x}_j(t); \theta_G), \\
    & \nabla_{\theta_G} b_i^{\text{mini}} = \frac{N}{S} \sum_{j \in B_2} \mathbf{A}_{ij} \nabla_{\theta_G} \mathbf{G}(\mathbf{x}_i(t), \mathbf{x}_j(t); \theta_G),
\end{aligned}
\]
where $S$ denotes the mini-batch size, and $\nabla_{\theta_G} \mathbf{G}(\mathbf{x}_i(t), \mathbf{x}_j(t); \theta_G)$ is a vector of size $K$.
\begin{proof}

The loss function for the $i$-th particle is given by
\begin{equation*}
    \ell_{i,t} = \left | \left( a_i - b_i^{\text{full}} \right) \right |^2.
\end{equation*}
Computing the gradient with respect to $\theta_G$, we obtain the gradient for the $i$-th particle,
\begin{equation*}
    \label{eqn:gradient}
    \nabla_{\theta_G} \ell_{i,t} = 2\left(a_i - b_i^{\text{full}}\right)\left(-\nabla_{\theta_G} b_i^{\text{full}}\right).
\end{equation*}
When using mini-batch gradient descent, we approximate the summation over $j$ by subsets $B_1, B_2$ of size $S$ chosen uniformly randomly from $N$ particles with two different random seeds. The mini-batch gradient for a single term then becomes
\begin{equation}
    \label{eqn:mini_batch_gradient}
    \nabla_{\theta_G} \ell_{i,t}^B = 2\left(a_i - b_i^{\text{mini}}\right) \left(-\nabla_{\theta_G}b_i^{\text{mini}}\right).
\end{equation}
Now, the mini-batch gradient is an unbiased estimator of the full gradient, i.e., $\mathbb{E}[\nabla_{\theta_G} \ell_{i,t}^B] = \nabla_{\theta_G} \ell_{i,t}$ by the independence of the two terms on the righ-hand side of Equation~\eqref{eqn:mini_batch_gradient}. Then, the noise term is:
\[
\begin{aligned}
Z_{i,t}(\theta_G) &= \nabla_{\theta_G} \ell_{i,t}^{B} - \nabla_{\theta_G} \ell_{i,t} \\
&= 2 \left( a_i - b_i^{\text{mini}} \right) \left( -\nabla_{\theta_G} b_i^{\text{mini}} \right) - 2 \left( a_i - b_i^{\text{full}} \right) \left( -\nabla_{\theta_G} b_i^{\text{full}} \right) \\
&= 2 \left( a_i \left( -\nabla_{\theta_G} b_i^{\text{mini}} \right) + b_i^{\text{mini}} \nabla_{\theta_G} b_i^{\text{mini}} + a_i \nabla_{\theta_G} b_i^{\text{full}} - b_i^{\text{full}} \nabla_{\theta_G} b_i^{\text{full}} \right) \\
&= 2 \left( a_i (-\nabla_{\theta_G} b_i^{\text{mini}} + \nabla_{\theta_G} b_i^{\text{full}}) + b_i^{\text{mini}} \nabla_{\theta_G} b_i^{\text{mini}} - b_i^{\text{full}} \nabla_{\theta_G} b_i^{\text{full}}\right).
\end{aligned}
\]
Using the triangle inequality, we get:
\begin{equation}
    \begin{aligned}
    \left\| b_i^{\text{full}\top} \nabla_{\theta_G} b_i^{\text{full}} - b_i^{\text{mini}\top} \nabla_{\theta_G} b_i^{\text{mini}} \right\|_2 &\leq \left\| b_i^{\text{full}\top} \nabla_{\theta_G} b_i^{\text{full}} - b_i^{\text{full}\top} \nabla_{\theta_G} b_i^{\text{mini}} \right\|_2 \\
    &\quad + \left\| b_i^{\text{full}\top} \nabla_{\theta_G} b_i^{\text{mini}} - b_i^{\text{mini}\top} \nabla_{\theta_G} b_i^{\text{mini}} \right\|_2.
    \end{aligned}
\end{equation}

Therefore, we have
\[
    \begin{aligned}
    \left \| Z_{i,t}(\theta_G) \right \|_2 &\leq 2 \left( \left| a_i \right| \left\| -\nabla_{\theta_G} b_i^{\text{mini}} + \nabla_{\theta_G} b_i^{\text{full}} \right\|_2 \right. \\
    &\quad + \left | b_i^{\text{full}} \right | \left\| \nabla_{\theta_G} b_i^{\text{full}} - \nabla_{\theta_G} b_i^{\text{mini}} \right\|_2 + \left. \left | b_i^{\text{full}} - b_i^{\text{mini}} \right| \left \| \nabla_{\theta_G} b_i^{\text{mini}} \right \|_2 \right).
    \end{aligned}
\]

And by Minkowski's inequality, we have
\[
    \begin{aligned}
    \mathbb{E}\left(\left\| Z_{i,t}(\theta_G) \right\|_2^q \right) &\leq 2^q \left( \mathbb{E}\left( | a_i |^q \left\| -\nabla_{\theta_G} b_i^{\text{mini}} + \nabla_{\theta_G} b_i^{\text{full}} \right\|_2^q \right)^{\frac{1}{q}} \right. \\
    &\quad + \mathbb{E}\left( \left| b_i^{\text{full}} \right|^q \left\| \nabla_{\theta_G} b_i^{\text{full}} - \nabla_{\theta_G} b_i^{\text{mini}} \right\|_2^q \right)^{\frac{1}{q}} \\
    &\quad + \left. \mathbb{E}\left( \left| b_i^{\text{full}} - b_i^{\text{mini}} \right|^q \| \nabla_{\theta_G} b_i^{\text{mini}} \|_2^q \right)^{\frac{1}{q}} \right)^q.
    \end{aligned}
\]

We consider bounding the term $\mathbb{E}\left|b_i^{\text{full}} - b_i^{\text{mini}} \right|^q $. By Hoeffding's theorem, we obtain
\[\mathbb{P}\left(\left|b_i^{\text{full}} - b_i^{\text{mini}}\right| \geq t \right) \leq 2\exp\left(-\frac{2St^2}{N^2 M_1^2} \right). \]

Then, 
\[ 
    \begin{aligned}
        \mathbb{E}\left(\left| b_i^{\text{mini}} - b_i^{\text{full}}\right|^q  \right) &= \int_{0}^{\infty} \mathbb{P}\left( \left|b_i^{\text{full}} - b_i^{\text{mini}} \right|^q > t\right) dt \\
        &= \int_{0}^{\infty} \mathbb{P}\left( \left|b_i^{\text{full}} - b_i^{\text{mini}} \right| > u\right) q u^{q-1} du \\
        &\leq \int_{0}^{\infty} 2\exp\left(-\frac{2Su^2}{N^2 M_1^2} \right) q u^{q-1} du \\
        &\leq \frac{q\left(NM_1\right)^q}{\left(2S\right)^{\frac{q}{2}}}\Gamma \left(\frac{q}{2}\right),
    \end{aligned}
\]
where $\Gamma(\cdot)$ is the gamma function. Then, $\mathbb{E}\left\| \nabla_{\theta_G} b_i^{\text{full}} - \nabla_{\theta_G} b_i^{\text{mini}} \right\|_2^q$ can be bounded similarly.
Using the fact that the event of the norm of difference of two vectors greater than $t$ is included in the event that at least one component of the difference is greater than $\frac{t}{\sqrt{K}}$, we have

\[
    \begin{aligned}
        \mathbb{P}\left( \left\| \nabla_{\theta_G} b_i^{\text{full}} - \nabla_{\theta_G} b_i^{\text{mini}} \right\|_2 > t \right) &\leq \sum_{k=1}^{K} \mathbb{P}\left( \left| \nabla_{\theta_G} b_i^{\text{full}, k} - \nabla_{\theta_G} b_i^{\text{mini}, k} \right| > \frac{t}{\sqrt{K}} \right) \\
        &\leq 2K \exp\left(-\frac{2S t^2}{N^2 M_2^2 K} \right).
    \end{aligned}
\]

Then, 
\[ 
    \begin{aligned}
        \mathbb{E}\left(\left\| \nabla_{\theta_G} b_i^{\text{full}} - \nabla_{\theta_G} b_i^{\text{mini}} \right\|_2^q \right) &= \int_{0}^{\infty} \mathbb{P}\left( \left\| \nabla_{\theta_G} b_i^{\text{full}} - \nabla_{\theta_G} b_i^{\text{mini}} \right\|_2^q > t\right) dt \\
        &= \int_{0}^{\infty} \mathbb{P}\left( \left\| \nabla_{\theta_G} b_i^{\text{full}} - \nabla_{\theta_G} b_i^{\text{mini}} \right\|_2 > u\right) q u^{q-1} du \\
        &\leq \int_{0}^{\infty} 2K\exp\left(-\frac{2Su^2}{N^2 M_2^2K} \right) q u^{q-1} du \\
        &\leq \frac{q\left(NM_2\right)^qK^{\frac{q}{2}+1}}{\left(2S\right)^{\frac{q}{2}}}\Gamma \left(\frac{q}{2}\right),
    \end{aligned}
\]

Combining all the bounds, we have
\[
    \mathbb{E}\left(\left\| Z_{i,t}(\theta_G) \right\|_2^q \right) \leq \frac{2^{\frac{q}{2}}q\Gamma(\frac{q}{2})\left(NM_2K^{\frac{1}{2}}\left(K^{\frac{1}{q}} + NM_1K^{\frac{1}{q}} + N  M_1\right)\right)^q}{S^{\frac{q}{2}}} .  
\]
Finally, we have

\[
\mathbb{E}\left[ \left\| \frac{1}{NT} \sum_{i=1}^N \sum_{t=1}^T Z_{i,t}(\theta_G) \right\|^q \right] \leq \frac{2^{\frac{q}{2}}q\Gamma(\frac{q}{2})N\left(M_2K^{\frac{1}{2}}\left(C_1K^{\frac{1}{q}} + NM_1K^{\frac{1}{q}} + N  M_1\right)\right)^q}{S^{\frac{q}{2}}T^{q-1}}.
\]
Then $\frac{1}{NT} \sum_{i=1}^N \sum_{t=1}^T Z_{i,t}(\theta_G)$ satisfies Assumption 4 in~\cite{mertikopoulos2020almost}, and by Theorem 4 in~\cite{mertikopoulos2020almost}, we conclude our proof.

\end{proof}

\section{Networks}\label{sec:networks}
In our work, we consider several synthetic networks to demonstrate the capabilities of FEX to identify dynamics on various network topologies.


\textbf{\text {Erd\H{o}s-R\'enyi} (ER) model}: ER model~\cite{van2009random,kang2014random} is one of the foundational models for random graphs. It is used widely to understand the properties of complex networks in various
fields, such as sociology, biology, and technology. In a ER model, a graph is constructed by connecting nodes randomly with a probability of $p$. We generate ER networks by using the $\textit{erdos\_renyi\_graph(n,p)}$ in $\textit{networkX}$~\cite{hagberg2008exploring}, where $n = 100, p = 0.05$. Here, $n$ denotes the number of nodes in the graph and $p$ represents the probability to connect two arbitrary nodes, respectively. To create a directed graph based on ER model, we allocate the direction of the link, i.e., from $i$ to $j$, from $j$ to $i$, or a reciprocal connection between them (both $i$ to $j$ and $j$ to $i$ directions exist), with an equal likelihood (each of probability $\frac{1}{3}$).

\textbf{Scale-Free (SF) network}: SF networks are ubiquitous in real-world applications~\cite{voitalov2019scale}. This type of network demonstrates an interesting phenomenon: nodes with more early connections tend to attract even more connections, leading to the ``rich get richer'' scenario. Therefore, the network's degree follows a power law distribution, implying a few nodes with very high degrees (hubs) and many with low degrees. Networks structured according to the scale-free model are constructed utilizing the Barab\'{a}si-Albert (BA)~\cite{albert2002statistical,barabasi1999emergence,barabasi2009scale} framework. Initially, we build such network by building a complete graph of a small number $m_0$ of nodes. Then at each time step, we add a new node with $m \leq m_0$ edges. The probability that the new node will connect to node $i$ is $P(i) = \frac{k_i}{
\sum_j k_j
}$, where $k_i$ is the degree of node $i$, and the sum is over all existing nodes $j$. Therefore, when the new node is added, it has a higher probability of connecting to nodes with high degrees, known as the ``preferential attachment''. We use the  $\textit{barabasi\_albert\_graph}$ function (with parameters $n = 100$, $m = 5$) in $\textit{NetworkX}$ to generate SF network, where $n$ is the number of nodes in the graph, and $m$ denotes each node connects to $m$ existing nodes. Initially, every link is established as bidirectional. Subsequently, a select portion of these unidirectional links are randomly removed.
An illustration of the ER model and the SF network is provided in Supplementary Fig.~\ref{fig:network}.

\begin{figure}[!ht]
    \centering
    \includegraphics[width=0.8\linewidth]{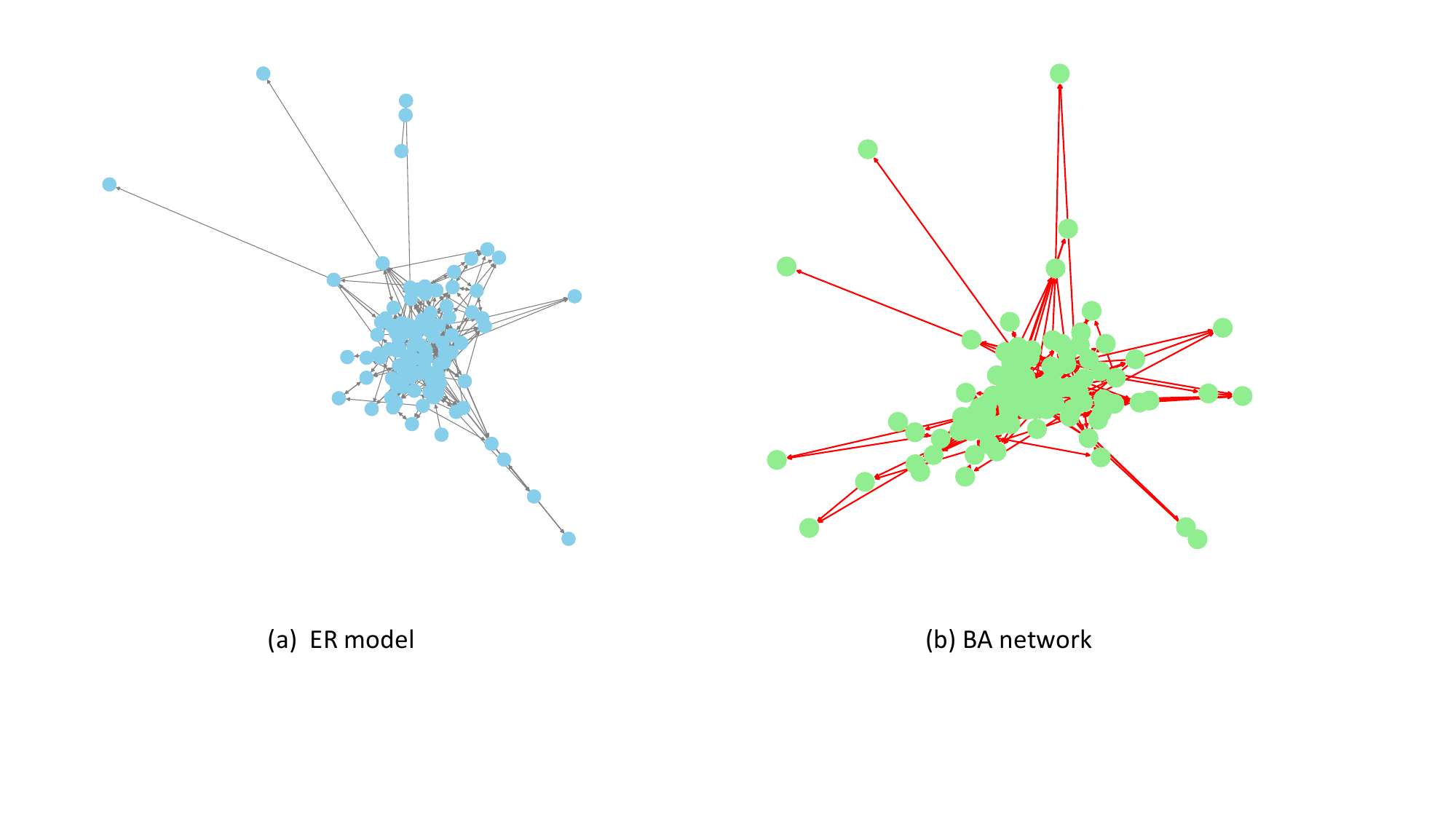}
    \caption{Visualization of ER model and SF (BA) network. (a) In ER network, the graph is essentially random without any structures or patterns. (b) In SF (BA) network, nodes closer to the boundary have generally low degrees, while there are some interior nodes with a high number of connections (hubs). }

    \label{fig:network}
\end{figure}

Despite being synthetic, the ER model and the SF network are classic and essential network models within the scientific community. Given that numerous real-world networks exhibit scale-free properties, the ability of FEX to identify physical laws on such networks accurately holds significant potential for applications in real-world data.

\section{Algorithm and experimental details } \label{alg:fex}
In this section, we provide an overview of the implementation details of FEX of identification of network dynamics. The training algorithm of FEX is summarized in Supplementary Algorithm~\ref{alg:workflow}. The details of hyperparameters in the algorithm are introduced as follows.

(1) \textit{Score computation}. The score of an operator sequence $\Be$ is updated with the Adam optimizer with a learning rate of 0.001 across $T_1$ iterations. Following this initial phase, refinement of the score is carried out through the BFGS optimizer with learning rate of 1 for up to 20 iterations.

(2) \textit{Operator sequence generation}. In FEX framework, a binary tree of variable depths is utilized, where the binary operator set is chosen as $\mathbbm{B}=\{+,-,\times\}$, and the unary operator set as $\mathbbm{U}=\{0, 1, \text{Id}, (\cdot)^2, (\cdot)^3, (\cdot)^4, \exp, \sin, \cos, \tanh, \text{sigmoid}\}$. The controller $\bm{\chi}_\Phi$ is parameterized as a neural network with parameter $\Phi$. It takes a constant input of dimension 20, while the output size is determined by the expression $n_1|\mathbbm{B}|+n_2|\mathbbm{U}|$. Within this expression, $n_1$ and $n_2$ denote the counts of binary and unary operators, respectively, and $|\cdot|$ denotes the cardinality of a set.

(3) \textit{Controller update}. The update of policy gradient  is implemented with a batch size of $M = 10$ over 300 iterations with the Adam optimizer at a learning rate of 0.002. To encourage the exploration of new operators, the $\epsilon$-greedy strategy is employed, wherein the probability of random sampling for an $\Be_i$ is fixed at 0.1.

(4) \textit{Candidate optimization}. The capacity of the candidate pool is denoted by $K = 15$. For every $\Be$ belonging to the set $\mathbb{P}$, the parameter $\theta$ is optimized via the Adam optimizer, initiated at a learning rate of 0.001 across 20,000 iterations. The learning rate follows a cosine decay schedule as in~\cite{he2019bag}. The number of sampling is chosen as $L = 5$, and each random sample contains $S = 20$ nodes.

\begin{algorithm}[!ht]  
    \caption{Fast FEX algorithm with RBM to identify dynamics on networks}  
    \label{alg:workflow} 
    \textbf{Input:} Adjacency matrix $\mathbf{A}$ of network; Time series data $\bx$ of time steps $T_s$ of $N$ nodes in the network; Loss functional $\mathcal{L}$; self-dynamics tree $\mathcal{T}_1$; Interaction dynamics tree $\mathcal{T}_2$; Searching loop iteration $T$; Coarse-tune iteration $T_1$ with Adam; Coarse-tune iteration $T_2$ with BFGS; Fine-tune iteration $T_3$ with Adam; Pool size $K$; Batch size $M$; Coefficient filtering threshold $\tau$; Number of sampling $L$; Sampling batch size $S$;
    
    \textbf{Output:} The solution $u(\bx;\mathcal{T}_1,\mathcal{T}_2,\hat{\Be},\hat{\Bf},\hat{\bm{\theta}}_1,\hat{\bm{\theta}}_2)$.
    \begin{algorithmic}[1]
        \State Initialize the agent $\bm{\chi}_{\phi,1}$ and $\bm{\chi}_{\phi,2}$ for tree $\mathcal{T}_1$, and $\mathcal{T}_2$, respectively
        \State $\mathbb{P} \gets \{\}$, 
        \For{$t$ from $1$ to $T$}
        \State Sample $M$ operator sequences $\{\Be^{(1)}, \Be^{(2)}, \cdots, \Be^{(M)}\}$ from $\bm{\chi}_{\phi,1}$
         \State Sample $M$ operator sequences $\{\Bf^{(1)}, \Bf^{(2)}, \cdots, \Bf^{(M)}\}$ from $\bm{\chi}_{\phi,2}$
        \For{$m$ from $1$ to $M$}
        \State Divide $\{1,2,\ldots,N\}$ into $n$ batches randomly
        \For{each batch $\mathcal{C}_q$}
            \State Compute the loss of particle $i$ ($i \in \mathcal{C}_q$) by 
                $$\frac{1}{NT_s}\sum_{i=1}^N\sum_{t =1}^{T_s}\left(\left|\mathcal{T}_{1}(\Be^m)(\bx_i,t) +\sum_{j \in \mathcal{C}_q}{\mathbf{A}_{ij}} \mathcal{T}_2(\Bf^m)(\bx_i,t) -\dot{\mathbf{x}}_i(t)\right|^2\right)  $$

        \EndFor
        \State  Optimize $\mathcal{L} (u(\bx;\mathcal{T}_1,\mathcal{T}_2,\Be^{(m)},\Bf^{(m)}, \bm{\theta}_1,\bm{\theta}_2))$ by coarse-tune with $T_1+T_2$ iterations.

        \State Compute the reward $R(\Be^{(m)},\Bf^{(m)})$ of $\Be^{(m)}$ and $\Bf^{(m)}$
        \If{$\Be^{(m)}$ and $\Bf^{(m)}$ belong to the top-$K$ of candidates}
        \State $\mathbb{P}$.append($\Be^{(m)}$ and $\Bf^{(m)}$)
        \State $\mathbb{P}$ pops some $\Be$ and $\Bf$ with the smallest reward when overloading
        \EndIf
        \EndFor
        \State Update $\bm{\chi}_{\phi,1}$ from Section 3.4.5
        \State Update $\bm{\chi}_{\phi,2}$ from Section 3.4.5
        \EndFor
            \For{$\Be$ and $\Bf$ in $\mathbb{P}$}  
            \For{$l$ from $1$ to $L$}
                \State Randomly sample $S$ nodes from the total $N$ nodes;
                \State Fine-tune $\mathcal{L} (u(\bx^S;\mathcal{T}_1,\mathcal{T}_2,\Be,\Bf,\bm{\theta}_1^S,\bm{\theta}_2^S))$ with $T_3$ iterations, apply coefficient filtering with threshold $\tau$.
            \EndFor
            \State average $\bm{\theta}_1^S,\bm{\theta}_2^S$ over $L$ runs to obtain $\hat{\bm{\theta}}_1,\hat{\bm{\theta}}_2$;

            \EndFor
        \State \Return the expression with the smallest fine-tune error. 
    \end{algorithmic}
\end{algorithm}

\section{Robustness analyses}\label{sec:analyses}

In practical situations, data are often of low resolution and noisy. To evaluate the robustness of our proposed method, we perform a series of tests under various conditions: low resolution data, spurious or missing connections within the graph structure. To mimic a low-resolution data scenario, we down-sample the time series of all nodes within the complex network regularly. A random set of edges within the graph is either inserted or removed to emulate situations with spurious or missing connections. This comprehensive testing framework aims to show that our method demonstrates robust performance across a range of realistic and noisy data environments. In the following subsections, we elaborate on each of the tests respectively. We use Two-Phase and ARNI as our baselines for inferring dynamics on complex network. Both of them require two comprehensive libraries, $L_F$ (self-dynamics) and $L_G$ (interaction dynamics). We summarize them in Supplementary Table~\ref{tab:self_dynamics} and Supplementary Table~\ref{tab:inter_dynamics}, respectively.

\begin{table}
    \centering
    \begin{tabular}{lc}
        \toprule
         Functions & $L_F$  \\
        \midrule
        Polynomial & $x_i, x_i^2, x_i^3, x_i^4$ \\
        \hline
        Exponential & $\exp{(x_i})$\\
        \hline
        Activation & $\sigma(x_i)$, $\tanh(x_i)$ \\
        \hline
        Rescaling & $\frac{x_i}{k_i^{\text{in}}}$ \\
        \hline
        Trigonometric & $\sin(x_i)$, $\cos{x_i}$ \\

        \bottomrule
    \end{tabular}
    \caption{Library functions (Two-Phase and ARNI) for self-dynamics.} 
    \label{tab:self_dynamics}
\end{table}

\begin{table}
    \centering
    \begin{tabular}{lllll}
        \toprule
         Functions & $L_G(x_j)$ & $L_G(x_i x_j)$ &  $L_G(x_j - x_i )$ & $x_i 
 L_G(x_j)$\\
        \midrule
        Polynomial & $x_j$, $x_j^2$ & $x_i x_j, (x_i x_j)^2$ & $x_j - x_i$, $(x_j - x_i)^2$ & $x_i x_j^2$ \\
        \hline
         Exponential & $\exp(x_j)$ & $\exp(x_i x_j)$ & $\exp(x_j - x_i)$ & $x_i \exp(x_j)$ \\
         \hline
         Activation & $\tanh(x_j)$ & $\tanh(x_i x_j)$ & $\tanh(x_j - x_i)$ & $x_i \tanh(x_j)$ \\
         
         & $\sigma(x_j)$ & $\sigma(x_i x_j)$ & $\sigma(x_j - x_i)$ & $x_i \sigma(x_j)$ \\
          \hline
         Rescaling & $\frac{x_j}{k_i^{\text{in}}}$ & $\frac{x_i x_j}{k_i^{\text{in}}}$ & $\frac{x_j - x_i}{k_i^{\text{in}}}$ & $\frac{x_i x_j}{k_i^{\text{in}}}$ \\
          \hline
         Trigonometric & $\sin(x_j)$ & $\sin(x_i x_j)$ & $\sin(x_j - x_i)$ & $x_i \sin(x_j)$ \\
          & $\cos(x_j)$ & $\cos(x_i x_j)$ & $\cos(x_j - x_i)$ & $x_i \cos(x_j)$ \\
        \bottomrule
    \end{tabular}
    \caption{Library functions (Two-Phase and ARNI) for interaction dynamics.} 
    \label{tab:inter_dynamics}
\end{table}

\subsection{Low resolution by experimental techniques}
In this subsection, we test the robustness of FEX with low resolution data. We test this on the Coupled R\"ossler dynamics on the directed SF network. To be more specific, we systematically perform down-sampling on the time series data of each node within the network, aiming to evaluate the impact of reduced data resolution on our methodology. Specifically, we apply two distinct down-sampling rates of $10\%$ and $5\%$. This process allows us to emulate scenarios of considerably lower data resolution, enabling an assessment of our method's performance and robustness under such constrained data conditions. 

The dynamics of the first dimension of Coupled R\"ossler dynamics identified by Two-Phase, ARNI and FEX are summarized in Supplementary Table~\ref{tab:ross_down_sample_0.10} and Supplementary Table~\ref{tab:ross_down_sample_0.05} with down-sampling rates $10\%$ and $5\%$, respectively. It can be observed that FEX consistently achieves better coefficient identification accuracies under different sampling rates, especially in the coefficients of interaction dynamics $\bG(x_i,x_j)$. The sMAPEs of all compared methods with different down-sampling rates are summarized in Supplementary Table~\ref{tab:mse_ross}.

\begin{table}
    \centering
    \begin{tabular}{lllll}
        \toprule
        & True Dynamics & Two-Phase &  ARNI & FEX \\
        \midrule
        $\mathbf{F}(x_i)$ &$-\omega_i x_{i,2} $ & $-1.0113 x_{i, 2}$ & $-1.0123 x_{i, 2}$ & $-1.0045 x_{i, 2}$  \\
        \hline
        &$-x_{i,3}$ & $-0.9783x_{i, 3}$ & $-0.9625x_{i, 3}$ & $-0.9991x_{i, 3}$  \\
        \hline
        $\mathbf{G}(x_i,x_j)$ &$0.15 (x_{j,1} - x_{i,1})$ & $0.0973 \left(x_{j, 1}-x_{i, 1}\right)$ & $0.1263 \left(x_{j, 1}-x_{i, 1}\right)$ & $0.1431 \left(x_{j, 1}-x_{i, 1}\right)$ \\

        \bottomrule
    \end{tabular}
    \caption{Numerical results for coupled R\"ossler dynamics (down-sampling rate $10\%$) in self-dynamics $\bF(x_i)$ and interaction dynamics $\bG(x_i, x_j)$ term by term.} 
    \label{tab:ross_down_sample_0.10}
\end{table}

\begin{table}
    \centering
    \begin{tabular}{lllll}
        \toprule
        & True Dynamics &  Two-Phase &  ARNI & FEX \\
        \midrule
        $\mathbf{F}(x_i)$ &$-\omega_i x_{i,2} $ & $-1.0114 x_{i, 2}$ & $-1.0275 x_{i, 2}$ & $-1.0063 x_{i, 2}$  \\
        \hline
       &$ -x_{i,3}$ & $-0.9251x_{i, 3}$ & $-0.9338x_{i, 3}$ & $-0.9741x_{i, 3}$  \\
        \hline
        $\mathbf{G}(x_i,x_j)$ &$0.15(x_{j,1} - x_{i,1})$ & $0.0826 \left(x_{j, 1}-x_{i, 1}\right)$ & $0.1045 \left(x_{j, 1}-x_{i, 1}\right)$ & $0.1348 \left(x_{j, 1}-x_{i, 1}\right)$ \\

        \bottomrule
    \end{tabular}
    \caption{Numerical results for coupled R\"ossler dynamics (down-sampling rate $5\%$) in self-dynamics $\bF(x_i)$ and interaction dynamics $\bG(x_i, x_j)$ term by term.}.
    \label{tab:ross_down_sample_0.05}
\end{table}

\begin{table}
    \centering
    \begin{tabular}{clll}
        \toprule
        Down-sampling rate & Two-Phase &  ARNI & FEX \\
        \midrule
         $10\%$ & 0.0765 & 0.0370 & 0.0087 \\
        \hline
        $5\%$ & 0.1114 & 0.0755 & 0.0232\\
        \bottomrule
    \end{tabular}
    \caption{sMAPE of Coupled R\"ossler dynamics with low resolution data.} 
    \label{tab:mse_ross}
\end{table}


\subsection{Spurious and missing links}

In real-world scenarios, accurately observing the topology of networks is often impractical, so a thorough evaluation of our method's robustness in the presence of spurious or missing links is necessary. Therefore, we select directed SF networks as our testbed, wherein we introduce perturbations by randomly adding or removing certain edges. Consequently, during offline phase, we generate the time series data with the true adjacency matrix of the SF network. During the online phase, we use perturbed network adjacency matrix to infer the underlying physical laws of the system. In particular, we randomly add or remove $15\%$ links in the underlying SF networks, respectively. The dynamics identified by Two-Phase, ARNI and FEX are summarized in Supplementary Table~\ref{tab:fhn_add_0.2} and Supplementary Table~\ref{tab:fhn_del_0.2}, respectively. We note that FEX exhibits robustness against graph perturbations, in scenarios of both spurious and missing links, attributed to its design which aims to limit the size of the generated expressions and the fine-tuning of parameters. The Two-Phase method is comparable to FEX, and shows decent robustness against graph topology perturbations, owing to its topological sampling in the second phase. On the contrary, ARNI falls short in performance on noisy network structures as it was initially proposed to infer network topology, thereby potentially leading to overfitting. ARNI infers many redundant terms in the dynamics with highly inaccurate coefficients of each term. The sMAPEs of all compared methods with $15\%$ spurious and missing links are summarized in Supplementary Table~\ref{tab:mse_fhn}.

\begin{table}
    \centering
    \begin{tabular}{lllll}
        \toprule
        & True &  Two-Phase &  ARNI & FEX \\
        \midrule
        $\mathbf{F}(x_i)$ &$x_{i,1} $ & $0.9897 x_{i, 1}$ & $6.8282x_{i,1}$ & $0.9807 x_{i, 1}$  \\
        \hline
       &$ -x_{i,1}^3$ & $-0.9733x_{i, 1}^3$ & $-0.1202x_{i, 1}^3$ & $-0.9979x_{i, 1}^3$  \\
        \hline
        &$ -x_{i,2}$ & $-0.9861x_{i, 2}$ & $-1.3184x_{i,2}$ & $-0.9975x_{i, 2}$  \\
        \hline
        & &  & $1.3564x_{i,1}^2$ &  \\
        \hline
        & &  & -88.3806 &  \\
        \hline        
        & &  & $-1.8841\exp(x_{i,1})$ &  \\
        \hline    
        
        $\mathbf{G}(x_i,x_j)$ &$-(x_{j,1} - x_{i,1})$ & $-0.9862\left(x_{j, 1}-x_{i, 1}\right)$ & $-22.69(x_{j,1} - x_{i,1})$ & $-1.015 \left(x_{j, 1}-x_{i, 1}\right)$ \\
        \hline
        & &  & $-18.4197\sin(x_{j,1} - x_{i,1})$ &  \\
        \hline
        & &  & $181.6294\sigma(x_{j,1} - x_{i,1})$ &  \\

        \bottomrule
    \end{tabular}
    \caption{Numerical results for FHN dynamics (randomly add $15\%$ edges) in self-dynamics $\bF(x_i)$ and interaction dynamics $\bG(x_i, x_j)$ term by term (each term of true or inferred $\bG(x_i, x_j)$ is understood to be divided by $k_{\text{in}}$, i.e., $\frac{\bG(x_i, x_j)}{k_{\text{in}}}$, where $k_{\text{in}}$ is the in-degree of node $i$).}.
    \label{tab:fhn_add_0.2}
\end{table}

\begin{table}
    \centering
    \begin{tabular}{lllll}
        \toprule
        & True &  Two-Phase &  ARNI & FEX \\
        \midrule
        $\mathbf{F}(x_i)$ &$x_{i,1} $ & $0.9828 x_{i, 1}$ & $0.6644x_{i,1}$ & $0.9835 x_{i, 1}$  \\
        \hline
       &$ -x_{i,1}^3$ & $-0.9910x_{i, 1}^3$ & $-0.5332x_{i, 1}^3$ & $-1.0038x_{i, 1}^3$  \\
        \hline
        &$ -x_{i,2}$ & $-0.9837x_{i, 2}$ & $-1.2174x_{i,2}$ & $-0.9916x_{i, 2}$  \\
        \hline
        & &  & $-0.2063\exp(x_{i,1})$ &  \\
        \hline
        & &  & 0.4520 &  \\
        \hline  
        & &  & $-0.2257x_{i,1}x_{i,2}$ &  \\
        \hline 
        & &  & $0.2303x_{i,1}^2x_{i,2}$ &  \\
        \hline 
        & &  & $-0.2061\exp(x_{i,1})$ &  \\
        \hline

        $\mathbf{G}(x_i,x_j)$ &$-(x_{j,1} - x_{i,1})$ & $-0.9812\left(x_{j, 1}-x_{i, 1}\right)$ & $-0.9477(x_{j,1} - x_{i,1})$ & $-1.0247 \left(x_{j, 1}-x_{i, 1}\right)$ \\
        \hline
        & &  & $-0.1134\sin(x_{j,1} - x_{i,1})$ &  \\
        \hline
        & &  & $0.6586\sigma(x_{j,1} - x_{i,1})$ &  \\
        \hline
        & &  & $0.4114x_{j,1}$ &  \\

        \bottomrule
    \end{tabular}
    \caption{Numerical results for FHN dynamics (randomly delete $15\%$ edges) in self-dynamics $\bF(x_i)$ and interaction dynamics $\bG(x_i, x_j)$ term by term (each term of true or inferred $\bG(x_i, x_j)$ is understood to be divided by $k_{\text{in}}$, i.e., $\frac{\bG(x_i, x_j)}{k_{\text{in}}}$, where $k_{\text{in}}$ is the in-degree of node $i$).}.
    \label{tab:fhn_del_0.2}
\end{table}

\begin{table}
    \centering
    \begin{tabular}{clll}
        \toprule
        Percent of edges perturbed & Two-Phase &  ARNI & FEX \\
        \midrule
         $15\%$ added & 0.0082 & 0.8437 & 0.0049 \\
         \hline
          $15\%$ deleted & 0.0077 & 0.7192 & 0.0067 \\
        \bottomrule
    \end{tabular}
    \caption{sMAPE of FHN dynamics with $15\%$ perturbed edges added or removed.} 
    \label{tab:mse_fhn}
\end{table}

\newpage
\bibliography{supp}
\bibliographystyle{plain}